\documentclass[preprint,preprintnumbers,amsmath,amssymb]{revtex4}
\usepackage{epsfig}
\usepackage{graphicx}% Include figure files
\usepackage{dcolumn}% Align table columns on decimal point
\usepackage{bm}% bold math
\usepackage{threeparttable}

\def\beq{\begin{equation}}
\def\eeq{\end{equation}}
\def\eeqn{\end{equation}}
\newcommand\iden{\leavevmode\hbox{\small1\normalsize\kern-.33em1}}

\newcommand{\bea} {\begin{eqnarray}}
\newcommand{\eea} {\end{eqnarray}}

\let\jnfont=\rm
\def\NPB#1,{{\jnfont Nucl.\ Phys.\ B }{\bf #1},}
\def\PLB#1,{{\jnfont Phys.\ Lett.\ B }{\bf #1},}
\def\EPJC#1,{{\jnfont Eur.\ Phys.\ Jour.\ C }{\bf #1},}
\def\PRD#1,{{\jnfont Phys.\ Rev.\ D }{\bf #1},}
\def\PRL#1,{{\jnfont Phys.\ Rev.\ Lett.\ }{\bf #1},}
\def\MPLA#1,{{\jnfont Mod.\ Phys.\ Lett.\ A }{\bf #1},}
\def\JPG#1,{{\jnfont J.\ Phys.\ G }{\bf #1},}
\def\CTP#1,{{\jnfont Commun.\ Theor.\ Phys.\ }{\bf #1},}
\def\JHEP#1,{{\jnfont JHEP \ }{\bf #1},}
\def\NPPS#1,{{\jnfont Nucl.\ Phys.\ Proc.\ Suppl.\ }{\bf #1},}
\def\CPC#1,{{\jnfont Comput.\ Phys.\ Commun.\ }{\bf #1},}
\def\CPL#1,{{\jnfont Chin.\ Phys.\ Lett. }{\bf #1},}
\def\APPB#1,{{\jnfont Acta\ Phys.\ Polon.\ B }{\bf #1},}
\def\PR#1,{{\jnfont Phys.\ Rept.\  }{\bf #1},}
\def\CHC#1,{{\jnfont Chin.\ Phys.\ C }{\bf #1},}

\def\lsim{\raise0.3ex\hbox{$<$\kern-0.75em\raise-1.1ex\hbox{$\sim$}}}
\def\gsim{\raise0.3ex\hbox{$>$\kern-0.75em\raise-1.1ex\hbox{$\sim$}}}

\begin{document}

\title{\ \\[10mm] Revisiting lepton-specific 2HDM in light of muon g-2 anomaly}

\author{Lei Wang$^{1}$, Jin Min Yang$^{2,3,4}$, Mengchao Zhang$^{5}$, Yang Zhang$^{6}$}
 \affiliation{$^1$ Department of Physics, Yantai University, Yantai
264005, P. R. China\\
$^2$ CAS Key Laboratory of Theoretical Physics, Institute of Theoretical Physics, Chinese Academy of Sciences, Beijing 100190, China\\
$^3$ School of Physical Sciences, University of Chinese Academy of Sciences, Beijing 100049, China\\
$^4$ Department of Physics, Tohoku University, Sendai 980-8578, Japan\\
$^5$ Center for Theoretical Physics of the Universe, Institute for Basic Science (IBS), Daejeon, 34126, Korea\\
$^6$ ARC Centre of Excellence for Particle Physics at the Tera-scale, School of Physics and Astronomy, Monash University, Melbourne, Victoria 3800, Australia
}

\preprint{CTPU-PTC-18-30}
\preprint{CoEPP-MN-18-8}

%---------------------------------------------------------------------------

\begin{abstract}
We examine the lepton-specific 2HDM as a solution of muon $g-2$ anomaly 
under various theoretical and experimental constraints,
especially the direct search limits from the LHC and the requirement of a strong first-order 
phase transition in the early universe. 
We find that the muon g-2 anomaly can be explained in the region of 32 $<\tan\beta<$ 80, 10 GeV $<m_A<$ 65 GeV,
260 GeV $<m_H<$ 620 GeV and 180 GeV $<m_{H^\pm}<$ 620 GeV after imposing the joint constraints from 
the theory, the precision electroweak data, the 125 GeV Higgs data, the leptonic/semi-hadronic $\tau$ decays,
the leptonic $Z$ decays and Br$(B_s \to \mu^+ \mu^-)$. 
The direct searches from the $h\to AA$ channels 
can impose stringent upper limits on Br$(h\to AA)$ and 
the multi-lepton event searches can 
sizably reduce the allowed region of $m_A$ and $\tan\beta$ (10 GeV $<m_A<$ 44 GeV 
and 32 $<\tan\beta<$ 60).
Finally, we find that the model can produce a strong first-order phase transition 
in the region of 14 GeV $<m_A<$ 25 GeV, 310 GeV $<m_H<$ 355 GeV and 250 GeV $<m_{H^\pm}<$ 295 GeV,
allowed by the explanation of the muon $g-2$ anomaly.
 
\end{abstract}

\maketitle

\section{Introduction}
The muon anomalous magnetic moment ($g-2$) is a
very precisely measured observable. The muon $g-2$ anomaly has been
a long-standing puzzle since the announcement by the E821 experiment
in 2001 \cite{mug21,mug22}. The experimental value has an approximate $3\sigma$
discrepancy from the SM prediction \cite{mu3sig1,mu3sig2,mu3sig3}. 
As a popular extension of the SM, the two Higgs doublet models (2HDM) have been 
applied to explain the muon $g-2$ anomaly in the literature 
\cite{mu2h1,mu2h1-1,mu2h2,mu2h3,mu2h4,mu2h5,mu2h6,mu2h7,mu2h8,mu2h9,mu2h10,mu2h11,mu2h12,mu2h13,mu2h14,mu2h15,
mu2h16,mu2h17,mu2h18,mu2h19,mu2h20,mu2h21,mu2h22,mu2h23,mu2h24,mu2h25}.
Among these extensions, the lepton-specific 2HDM (L2HDM) provides a simple explanation for
the muon $g-2$ anomaly \cite{mu2h5,mu2h8,mu2h9,mu2h10,mu2h11,mu2h16}.
A light pseudoscalar with a large coupling to lepton can sizably enhance the muon $g-2$ via the two-loop Barr-Zee diagrams.

After the discovery of the SM-like Higgs boson at the LHC, it was found \cite{mu2h9} that
the muon $g-2$ explanation favors the lepton Yukawa couplings of the SM-like Higgs to have  
an opposite sign with respect to the SM couplings.
The observation of Br$(B_s \to \mu^+ \mu^-)$ gives a new 
constraint on the parameter space of L2HDM \cite{mu2h9}. 
Further, it was found \cite{mu2h10}  that the leptonic
$Z$ decays and leptonic/semi-hadronic $\tau$ decays can also give strong constraints on the parameter space
of L2HDM, and a more precise calculation is performed in \cite{mu2h16}.
The L2HDM can lead to $\tau-$rich signatures at the LHC in the parameter 
region favored by muon $g-2$. The study in \cite{mu2h11} derived the constraints on the model 
by using the chargino/neutralino search at the 8 Tev LHC and analyzed the prospects at the 14 TeV LHC.

In this work we examine the parameter space of L2HDM by considering the
joint constraints from the theory, the precision electroweak data,
 the 125 GeV Higgs signal data, the muon $g-2$ anomaly, the lepton flavor universality (LFU) in the $\tau$ and $Z$ decays, 
the measurement of Br$(B_s \to \mu^+ \mu^-)$, as well as the direct search limits from the LHC (the data of some channels 
analyzed by the ATLAS and CMS are corresponding to an integrated luminosity up to about 36 $fb^{-1}$ recorded in
proton-proton collisions at $\sqrt{s}$ = 13 TeV). 
On the other hand, it is known that the 2HDM can trigger a strong first-order phase transition (SFOPT) 
in the early universe~\cite{PT_2HDM,Cline:2011mm}, 
which is required by a successful explanation of the observed baryon asymmetry of the universe (BAU)~\cite{Sakharov:1967dj} 
and can produce primordial gravitational-wave (GW) signals~\cite{PT_GW} potentially detectable 
by future space-based laser interferometer detectors like eLISA~\cite{eLISA}.
Due to the importance of SFOPT in cosmology, we will also analyze whether a SFOPT is achievable in the  
parameter space in favor of the muon $g-2$ explanation. 

Our work is organized as follows. In Sec. II we recapitulate the
L2HDM. In Sec. III we discuss the muon $g-2$ anomaly and other relevant constraints. In Sec.
IV, we constrain the model using the direct search limits from the LHC. 
In Sec. V, we discuss some benchmark scenarios leading to a SFOPT.
Finally, we give our conclusion in Sec. VI.

\section{The lepton-specific 2HDM}
The Higgs potential with a softly-broken discrete $Z_2$ symmetry  is given as \cite{2h-poten}
\begin{eqnarray} \label{V2HDM} \mathrm{V} &=& m_{11}^2
(\Phi_1^{\dagger} \Phi_1) + m_{22}^2 (\Phi_2^{\dagger}
\Phi_2) - \left[m_{12}^2 (\Phi_1^{\dagger} \Phi_2 + \rm h.c.)\right]+ \frac{\lambda_1}{2}  (\Phi_1^{\dagger} \Phi_1)^2\nonumber \\
&& +\frac{\lambda_2}{2} (\Phi_2^{\dagger} \Phi_2)^2 + \lambda_3
(\Phi_1^{\dagger} \Phi_1)(\Phi_2^{\dagger} \Phi_2) + \lambda_4
(\Phi_1^{\dagger}
\Phi_2)(\Phi_2^{\dagger} \Phi_1)+ \left[\frac{\lambda_5}{2} (\Phi_1^{\dagger} \Phi_2)^2 + \rm
h.c.\right].
\end{eqnarray}
In this paper we focus on the CP-conserving case where all
$\lambda_i$ and $m_{12}^2$ are real. The two complex
scalar doublets respectively have the vacuum expectation values (VEVs) $v_1$ and $v_2$ 
with $v^2 = v^2_1 + v^2_2 =
(246~\rm GeV)^2$, and the ratio of the two VEVs is defined as usual
to be $\tan\beta=v_2 /v_1$. There are five mass eigenstates: two
neutral CP-even states $h$ and $H$, one neutral pseudoscalar $A$, and two
charged scalars $H^{\pm}$. 

In the L2HDM, the quarks obtain masses from $\Phi_2$ field, and the leptons from
$\Phi_1$ field \cite{xy-1,xy-3}. The Yukawa interactions are given by
 \bea
- {\cal L} &=&Y_{u2}\,\overline{Q}_L \, \tilde{{ \Phi}}_2 \,u_R
+\,Y_{d2}\,\overline{Q}_L \, { \Phi}_2 \,d_R + \, Y_{\ell 1}\,\overline{L}_L \, {\Phi}_1\,e_R+\, \mbox{h.c.}\,, \eea where
$Q_L^T=(u_L\,,d_L)$, $L_L^T=(\nu_L\,,l_L)$,
$\widetilde\Phi_{1,2}=i\tau_2 \Phi_{1,2}^*$, and $Y_{u2}$,
$Y_{d2}$ and $Y_{\ell 1}$ are $3 \times 3$ matrices in family
space.

The Yukawa couplings of the neutral Higgs bosons normalized to the SM are given by
\bea\label{hffcoupling} &&
y^{h}_V=\sin(\beta-\alpha),~~~y_{f}^{h}=\left[\sin(\beta-\alpha)+\cos(\beta-\alpha)\kappa_f\right], \nonumber\\
&&y^{H}_V=\cos(\beta-\alpha),~~~y_{f}^{H}=\left[\cos(\beta-\alpha)-\sin(\beta-\alpha)\kappa_f\right], \nonumber\\
&&y^{A}_V=0,~~~y_{A}^{f}=-i\kappa_f~{\rm (for}~u),~~~~y_{f}^{A}=i \kappa_f~{\rm (for}~d,~\ell),
\eea
where  $V$ denotes $Z$ or $W$,  
$\kappa_\ell\equiv-\tan\beta$, $\kappa_d=\kappa_u\equiv 1/\tan\beta$ and  
$\alpha$ is the mixing angle of the two CP-even Higgs bosons.

\section{Muon $g-2$ anomaly and relevant constraints}\label{constraints}
\subsection{Numerical calculations}
In this paper, the light CP-even Higgs $h$ is taken as the SM-like Higgs, $m_h=$ 125 GeV. We take 
a convention \cite{2hc-1}, $0\leq\beta\leq \frac{\pi}{2}$ and $-\frac{\pi}{2}\leq\beta-\alpha\leq\frac{\pi}{2}$,
which leads to $0\leq\cos(\beta-\alpha)\leq 1$ and $-1\leq\sin(\beta-\alpha)\leq 1$.
We take $\tan\beta$ and $\sin(\beta-\alpha)$ as input parameters, which replace the mixing angles
$\beta$ and $\alpha$, respectively. If $\sin(\beta-\alpha)$ and $\tan\beta$ are
given, we can determine $\beta$ and $\alpha$ by $\beta=\arctan\beta$ and 
$\alpha=\arctan\beta-\arcsin(\beta-\alpha)$, respectively.
Since the muon $g-2$ anomaly favors
a light pseudoscalar with a large coupling to lepton, we scan over $m_A$ and $\tan\beta$ in the following
ranges:
\beq
10 ~{\rm GeV} <m_A< 120~ {\rm GeV},~~20<\tan\beta<120.
\eeq
Such $\tan\beta$ can make $\mid y_f^h\mid$ to deviate from 1 sizably for a large $\cos(\beta-\alpha)$, 
which is disfavored by the signal data of the 125 GeV Higgs. Therefore, we take $\mid\sin(\beta-\alpha)\mid$
to be close to 1. According to the results on $\sin(\beta-\alpha)$ in Ref. \cite{mu2h9}, we scan over $\sin(\beta-\alpha)$
in the following ranges:
\beq
0.994\leq\sin(\beta-\alpha)\leq 1,~~~ -1\leq\sin(\beta-\alpha)\leq-0.994.
\eeq

In our calculation, we consider the following observables and constraints:

\begin{itemize}
\item[(1)] Theoretical constraints and precision electroweak data. The $\textsf{2HDMC}$ \cite{2hc-1}
is employed to implement the theoretical
constraints from the vacuum stability, unitarity and
coupling-constant perturbativity, as well as the constraints from
the oblique parameters ($S$, $T$, $U$).

\item[(2)] The signal data of the 125 GeV Higgs. 
Since the 125 GeV Higgs couplings with the SM particles in this model
can deviate from the SM ones, the SM-like decay modes will be modified. 
Besides, for $m_A$ is smaller than 62.5 GeV, the invisible decay $h\to AA$ is kinematically allowed,
which will be strongly constrained by the experimental data of the 125 GeV Higgs. 
 We perform $\chi^2_h$ calculation for the signal strengths of the 125 GeV Higgs in the
$\mu_{ggF+tth}(Y)$ and $\mu_{VBF+Vh}(Y)$ with $Y$ denoting the decay
mode $\gamma\gamma$, $ZZ$, $WW$, $\tau^+ \tau^-$ and $b\bar{b}$,
 \begin{eqnarray} \label{eq:ellipse}
  \chi^2(Y) =\left( \begin{array}{c}
        \mu_{ggH+ttH}(Y) - \widehat{\mu}_{ggH+ttH}(Y)\\
        \mu_{VBF+VH}(Y) - \widehat{\mu}_{VBF+VH}(Y)
                 \end{array} \right)^T
                 \left(\begin{array}{c c}
                        a_Y & b_Y \\
                        b_Y & c_Y
                 \end{array}\right) \nonumber\\
\times
                  \left( \begin{array}{c}
        \mu_{ggH+ttH}(Y) - \widehat{\mu}_{ggH+ttH}(Y)\\
        \mu_{VBF+VH}(Y) - \widehat{\mu}_{VBF+VH}(Y)
                 \end{array} \right) \,.
 \end{eqnarray}
where $\widehat{\mu}_{ggH+ttH}(Y)$ and $\widehat{\mu}_{VBF+VH}(Y)$
are the data best-fit values and $a_Y$, $b_Y$ and $c_Y$ are the
parameters of the ellipse, which are given by the
combined ATLAS and CMS experiments \cite{160602266}.

\item[(3)] LFU in the $\tau$ decays.
The HFAG collaboration reported three ratios from pure leptonic processes, and two ratios
from semi-hadronic processes, $\tau \to \pi/K \nu$ and $\pi/K \to \mu \nu$ \cite{tauexp}: 
\begin{eqnarray} \label{hfag-data}
&&
\left( g_\tau \over g_\mu \right) =1.0011 \pm 0.0015,~~
\left( g_\tau \over g_e \right) = 1.0029 \pm 0.0015,~~ 
\left( g_\mu \over g_e \right) = 1.0018 \pm 0.0014, 
\nonumber\\
&&
\left( g_\tau \over g_\mu \right)_\pi = 0.9963 \pm 0.0027, \quad
\left( g_\tau \over g_\mu \right)_K = 0.9858 \pm 0.0071,
\end{eqnarray}
The correlation matrix for the above five observables is
\begin{equation} \label{hfag-corr}
\left(
\begin{array}{ccccc}
1 & +0.53 & -0.49 & +0.24 & +0.12 \\
+0.53  & 1     &  + 0.48 & +0.26    & +0.10 \\
-0.49  & +0.48  & 1       &   +0.02 & -0.02 \\
+0.24  & +0.26  & +0.02  &     1    &     +0.05 \\
+0.12  & +0.10  & -0.02  &  +0.05  &   1 
\end{array} \right) .
\end{equation}
In the L2HDM we have the ratios
\begin{eqnarray} \label{deltas-data}
&&
\left( g_\tau \over g_\mu \right) \approx 1+ \delta_{\rm loop}, \quad
\left( g_\tau \over g_e \right) \approx 1+ \delta_{\rm tree}+ \delta_{\rm loop}, \quad
\left( g_\mu \over g_e \right) \approx 1+ \delta_{\rm tree}, 
\nonumber\\
&&
\left( g_\tau \over g_\mu \right)_\pi \approx 1+ \delta_{\rm loop}, \quad
\left( g_\tau \over g_\mu \right)_K \approx 1+ \delta_{\rm loop} ,
\end{eqnarray}
where $\delta_{\rm tree}$ and $\delta_{\rm loop}$ are respectively corrections from
the tree-level diagrams and the one-loop diagrams mediated by the charged Higgs. They are given as \cite{mu2h10,mu2h16}
\begin{eqnarray} \label{deltas}
\delta_{\rm tree} &=& {m_\tau^2 m_\mu^2 \over 8 m^4_{H^\pm}} t^4_\beta
- {m_\mu^2 \over m^2_{H^\pm}} t^2_\beta {g(m_\mu^2/m^2_\tau) \over f(m_\mu^2/m_\tau^2)}, \\
\delta_{\rm loop} &=& {1 \over 16 \pi^2} { m_\tau^2 \over v^2}  t^2_\beta
\left[1 + {1\over4} \left( H(x_A) + s^2_{\beta-\alpha} H(x_H) + c^2_{\beta-\alpha} H(x_h)\right)
\right]\,, 
\end{eqnarray}
where $f(x)\equiv 1-8x+8x^3-x^4-12x^2 \ln(x)$, $g(x)\equiv 1+9x-9x^2-x^3+6x(1+x)\ln(x)$ and
$H(x_\phi) \equiv \ln(x_\phi) (1+x_\phi)/(1-x_\phi)$ with $x_\phi=m_\phi^2/m_{H^{\pm}}^2$.

We perform $\chi^2_\tau$ calculation for the five observables. The covariance matrix constructed from the data of Eq. (\ref{hfag-data})
and Eq. (\ref{hfag-corr}) has a vanishing eigenvalue, and the corresponding degree is removed in our calculation.

\item[(4)] LFU in the $Z$ decays. The measured values of the ratios of the leptonic $Z$ decay
branching fractions are given as \cite{zexp}:
\begin{eqnarray} \label{lu-zdecay}
{\Gamma_{Z\to \mu^+ \mu^-}\over \Gamma_{Z\to e^+ e^- }} &=& 1.0009 \pm 0.0028
\,,\\ 
{\Gamma_{Z\to \tau^+ \tau^- }\over \Gamma_{Z\to e^+ e^- }} &=& 1.0019 \pm 0.0032
\,,
\end{eqnarray}
with a correlation of $+0.63$. 
In the L2HDM, the width of $Z\to \tau^+\tau^-$ can have sizable deviation from the SM value
by the loop contributions of the extra Higgs bosons, because they strongly interact with charged
leptons for large $\tan\beta$. The quantities of Eq. (\ref{lu-zdecay}) are calculated in the L2HDM
 as \cite{mu2h10,mu2h16}
\begin{eqnarray} 
&&{\Gamma_{Z\to \mu^+ \mu^-}\over \Gamma_{Z\to e^+ e^- }} \approx 1.0+ {2 g_L^e{\rm Re}(\delta g^{\rm 2HDM}_L)+ 2 g_R^e{\rm Re}(\delta g^{\rm 2HDM}_R) \over {g_L^e}^2 + {g_R^e}^2 }\frac{m_\mu^2}{m_\tau^2}\,.
\,,\\ 
&&{\Gamma_{Z\to \tau^+ \tau^- }\over \Gamma_{Z\to e^+ e^- }} 
\approx 1.0+ {2 g_L^e{\rm Re}(\delta g^{\rm 2HDM}_L)+ 2 g_R^e{\rm Re}(\delta g^{\rm 2HDM}_R) \over {g_L^e}^2 + {g_R^e}^2 }\,.
\,
\end{eqnarray}
where the SM value $g_L^e=-0.27$ and $g_R^e=0.23$. $\delta g^{\rm 2HDM}_L$ and $\delta g^{\rm 2HDM}_R$
are from the one-loop corrections of L2HDM, which are explicitly given in Ref. \cite{mu2h16}.

\item[(5)] The muon $g-2$. The recent measurement is 
$a_\mu^{exp}=(116592091\pm63)\times10^{-11}$
\cite{muonexp}, which has approximately 3.1$\sigma$ deviation from
the SM prediction \cite{muonsm}, $\Delta
a_\mu=a_\mu^{exp}-a_\mu^{SM}=(262\pm85)\times10^{-11}.$
In this paper, we require the model to explain the muon $g-2$ anomaly at the $2\sigma$ level.

In the L2HDM, the muon $g-2$ obtains contributions from the
one-loop diagrams induced by the Higgs bosons and also from the
two-loop Barr-Zee diagrams mediated by $A$, $h$ and $H$. For the
one-loop contributions \cite{mu2h1} we have
 \beq
    \Delta a_\mu^{\mbox{$\scriptscriptstyle{\rm 2HDM}$}}({\rm 1loop}) =
    \frac{G_F \, m_{\mu}^2}{4 \pi^2 \sqrt{2}} \, \sum_j
    \left (y_{\mu}^j \right)^2  r_{\mu}^j \, f_j(r_{\mu}^j),
\label{amuoneloop}
\end{equation}
where $j = h,~ H,~ A ,~ H^\pm$, $r_{\mu}^ j =  m_\mu^2/M_j^2$. For
$r_{\mu}^j\ll$ 1 we have
\beq
    f_{h,H}(r) \simeq- \ln r - 7/6,~~
    f_A (r) \simeq \ln r +11/6, ~~
    f_{H^\pm} (r) \simeq -1/6.
    \label{oneloopintegralsapprox3}
\eeq
The two-loop contributions are given by  
\beq
    \Delta a_\mu^{\mbox{$\scriptscriptstyle{\rm 2HDM}$}}({\rm 2loop-BZ})
    = \frac{G_F \, m_{\mu}^2}{4 \pi^2 \sqrt{2}} \, \frac{\alpha_{\rm em}}{\pi}
    \, \sum_{i,f}  N^c_f  \, Q_f^2  \,  y_{\mu}^i  \, y_{f}^i \,  r_{f}^i \,  g_i(r_{f}^i),
\label{barr-zee}
\end{equation}
where $i = h,~ H,~ A$, and $m_f$, $Q_f$ and $N^c_f$ are the mass,
electric charge and the number of color degrees of freedom of the
fermion $f$ in the loop. The functions $g_i(r)$ are \cite{mu2h1-1,mu2h2,mu2h4}
\begin{eqnarray}
    && g_{h,H}(r) = \int_0^1 \! dx \, \frac{2x (1-x)-1}{x(1-x)-r} \ln
    \frac{x(1-x)}{r}, \\
    && g_{A}(r) = \int_0^1 \! dx \, \frac{1}{x(1-x)-r} \ln
    \frac{x(1-x)}{r}.
\end{eqnarray}
The contributions of the CP-even (CP-odd) Higgses to $a_\mu$
are negative (positive) at the two-loop level and positive
(negative) at one-loop level. As $m^2_f/m^2_\mu$ could easily
overcome the loop suppression factor $\alpha/\pi$, the two-loop
contributions can be larger than one-loop ones.

\item[(6)] $B_s\to \mu^+\mu^-$. We take the formulas in \cite{li-bsuu} to calculate 
$B_s\to \mu^+\mu^-$
\begin{equation} \label{rbsuu}
 \frac{\overline{\mathcal{B}}(B_s\to \mu^+\mu^-)}
 {\overline{\mathcal{B}}(B_s\to \mu^+\mu^-)_{\rm SM}}
\, =\,
\bigg[\,|P|^2+\Big(1-\frac{\Delta\Gamma_s}{\Gamma^s_L}\,\Big)|S|^2\bigg]\,,
\end{equation}
where the CKM matrix elements and hadronic factors cancel out, and 
\begin{align}
 P &\equiv\, \frac{C_{10}}{C^{\rm SM}_{10}} + \frac{M^2_{B_s}}{2M^2_W}
 \left(\frac{m_b}{m_b+m_s}\right)\,\frac{C_P-C_P^{\mathrm{SM}}}{C^{\rm
 SM}_{10}},
  \,
 \label{eq:P}\\[0.2cm]
 S &\equiv\, \sqrt{1-\frac{4m^2_\mu}{M^2_{B_s}}}\; \frac{M^2_{B_s}}{2M^2_W}
 \left(\frac{m_b}{m_b+m_s}\right)\,\frac{C_S-C_S^{\mathrm{SM}}}{C^{\rm SM}_{10}}
 \,.
 \label{eq:S}
\end{align}
The L2HDM can give the additional contributions to coefficient
$C_{10}$ by the $Z$-penguin diagrams with the charged Higgs loop.
Unless there are large enhancements for $C_P$ and $C_S$, their
contributions can be neglected due to the suppression of the factor
$M_{B_s}^2/M_W^2$. In the L2HDM, $C_P$ can obtain the important contributions from the
CP-odd Higgs exchange diagrams for a very small $m_A$. 
The experimental data of Br$(B_s\to \mu\mu)$ is given as \cite{expbsuu}
\beq
Br(B_s\to \mu\mu)=(3.0\pm0.6^{+0.3}_{-0.2})\times 10^{-9}.
\eeq

\begin{table}
\begin{footnotesize}
\begin{tabular}{| c | c | c | c |}
\hline
\textbf{Channel} & \textbf{Experiment} & \textbf{Mass range (GeV)}  &  \textbf{Luminosity} \\
\hline
%%%%%%%%%%%%
 {$gg\to h \to AA \to \tau^{+}\tau^{-}\tau^{+}\tau^{-}$} & ATLAS 8 TeV~\cite{1505.01609} & 4-50 & 20.3 fb$^{-1}$ \\
{$pp\to  h \to AA \to \tau^{+}\tau^{-}\tau^{+}\tau^{-}$} & CMS 8 TeV~\cite{1701.02032} &  5-15  &19.7 fb$^{-1}$ \\
{$pp\to  h \to AA \to (\mu^{+}\mu^{-})(b\bar{b})$} & CMS 8 TeV~\cite{1701.02032} &  25-62.5  &19.7 fb$^{-1}$ \\
{$pp\to  h \to AA \to (\mu^{+}\mu^{-})(\tau^{+}\tau^{-})$} & CMS 8 TeV~\cite{1701.02032} &  15-62.5  &19.7 fb$^{-1}$ \\
\hline

\end{tabular}
\end{footnotesize}
\caption{The upper limits at 95\%  C.L. on the production cross-section times branching ratio for $h\to AA$ channels
at the LHC.}
\label{tabh}
\end{table}

\item[(7)] The exclusion limits from the searches for Higgs bosons at the LEP 
and $h\to AA$ at the LHC.
We employ $\textsf{HiggsBounds}$ \cite{hb1,hb2} to implement the exclusion
constraints from the searches for the neutral and charged Higgs at the LEP 
at 95\% confidence level.
The searches for a light Higgs at the LEP can impose stringent constraints
on the parameter space.

The ATLAS and CMS have searched for some exotic decay channels of the 125 GeV Higgs, such as 
$h\to AA$. In addition to the global fit to the 125 GeV Higgs signal data, the $hAA$ coupling will
be constrained by the ATLAS and CMS direct searches for $h\to AA$ channels at the LHC.
Table \ref{tabh} shows several $h\to AA$ channels considered by us.

\end{itemize}

The 125 GeV Higgs signal data and the LFU data from $\tau$ decays include a large number of observales.
We perform a global fit to the 125 GeV Higgs signal data and the LFU data from $\tau$ decays, and 
define $\chi^2$ as $\chi^2=\chi^2_h+\chi^2_\tau$.
 We pay particular attention to the surviving samples with
$\chi^2-\chi^2_{\rm min} \leq 6.18$, where $\chi^2_{\rm min}$
denotes the minimum of $\chi^2$. These samples correspond to be within
the $2\sigma$ range in any two-dimension plane of the
model parameters when explaining the signal data of the 125 GeV Higgs and the data of the LFU from $\tau$ decays.

%%%%%%%%%%%%%%%%%%%%%
\begin{figure}[tb]
%\begin{center}
  \epsfig{file=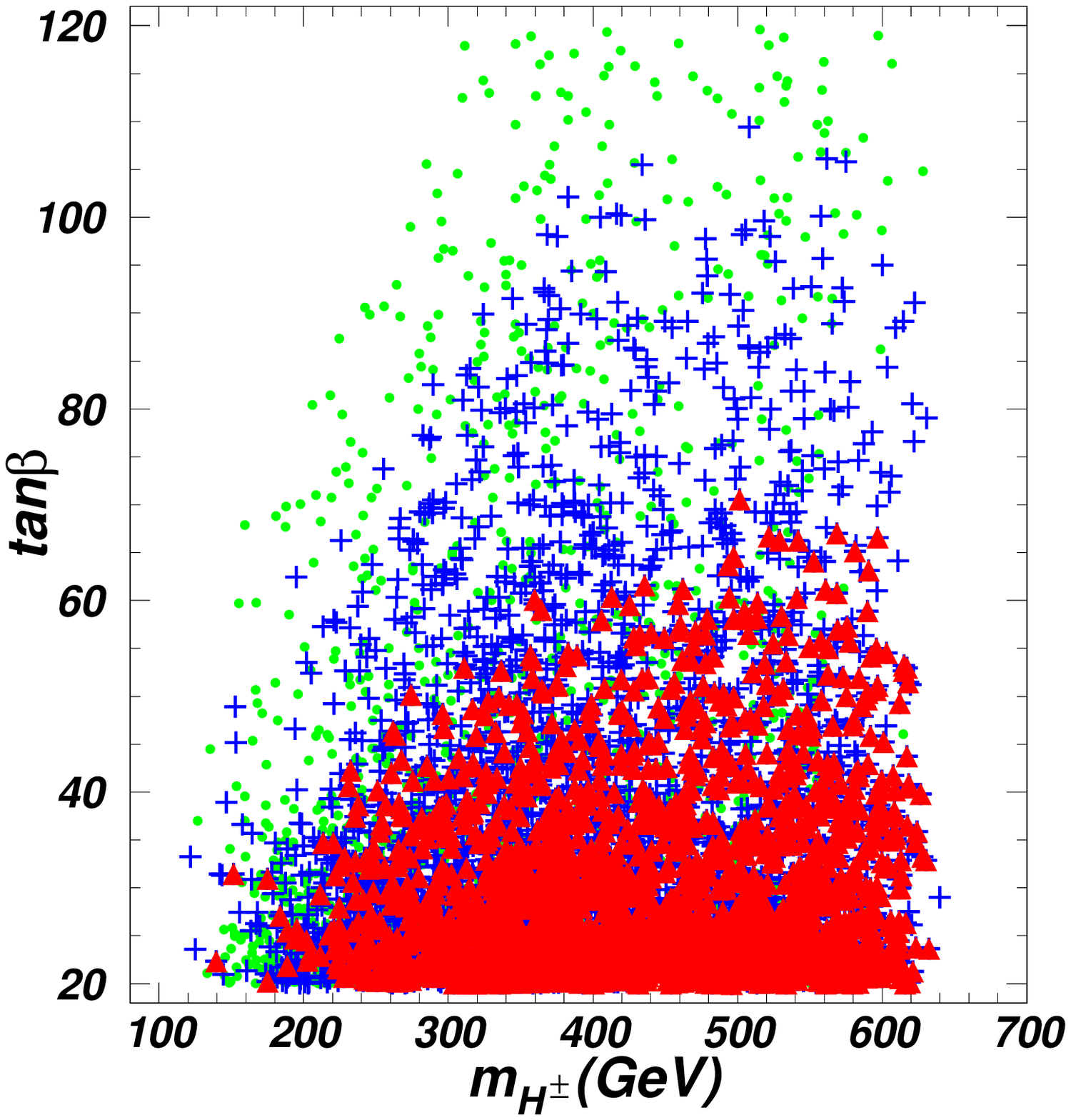,height=7.3cm}
  \epsfig{file=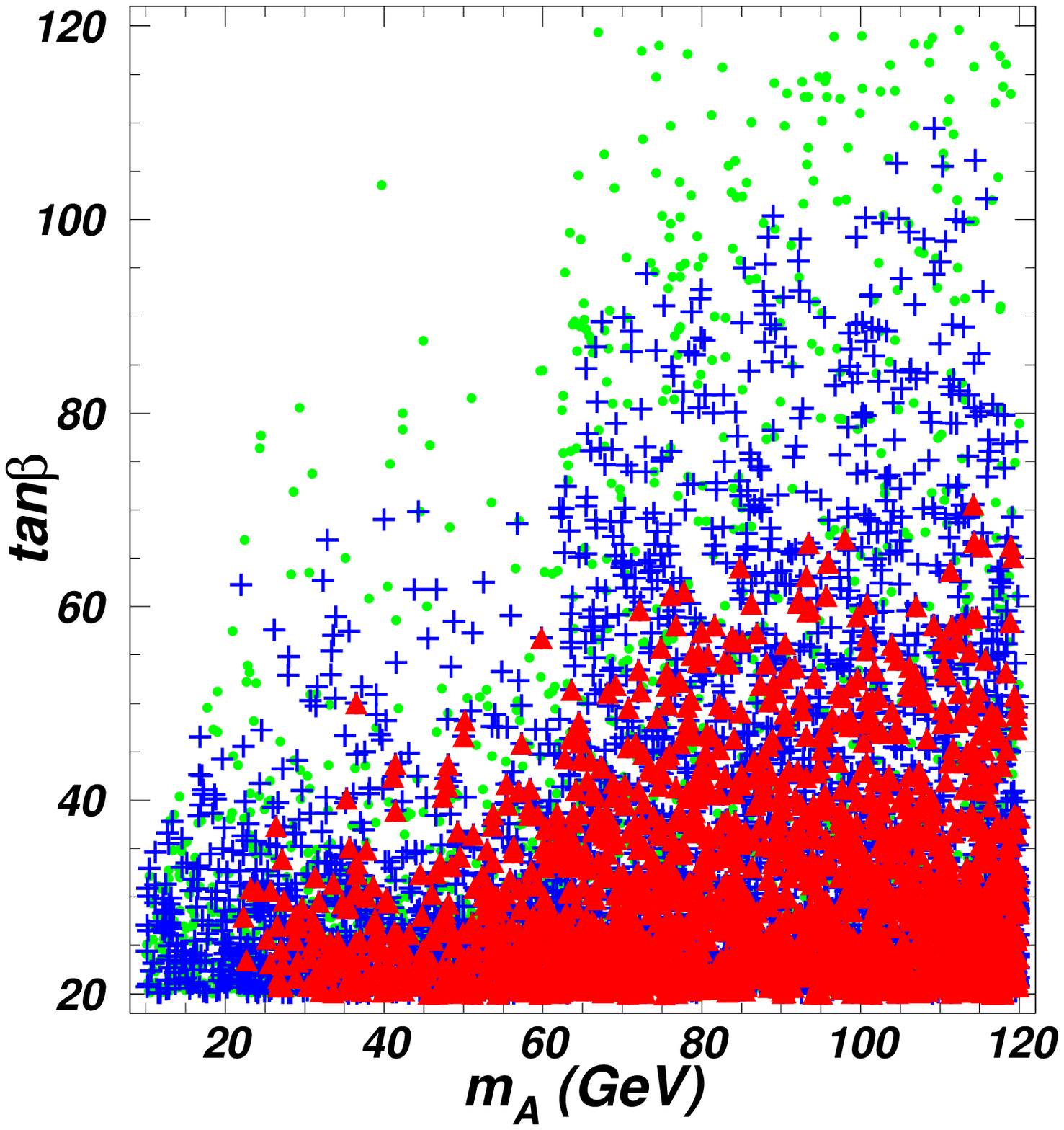,height=7.3cm}
 \epsfig{file=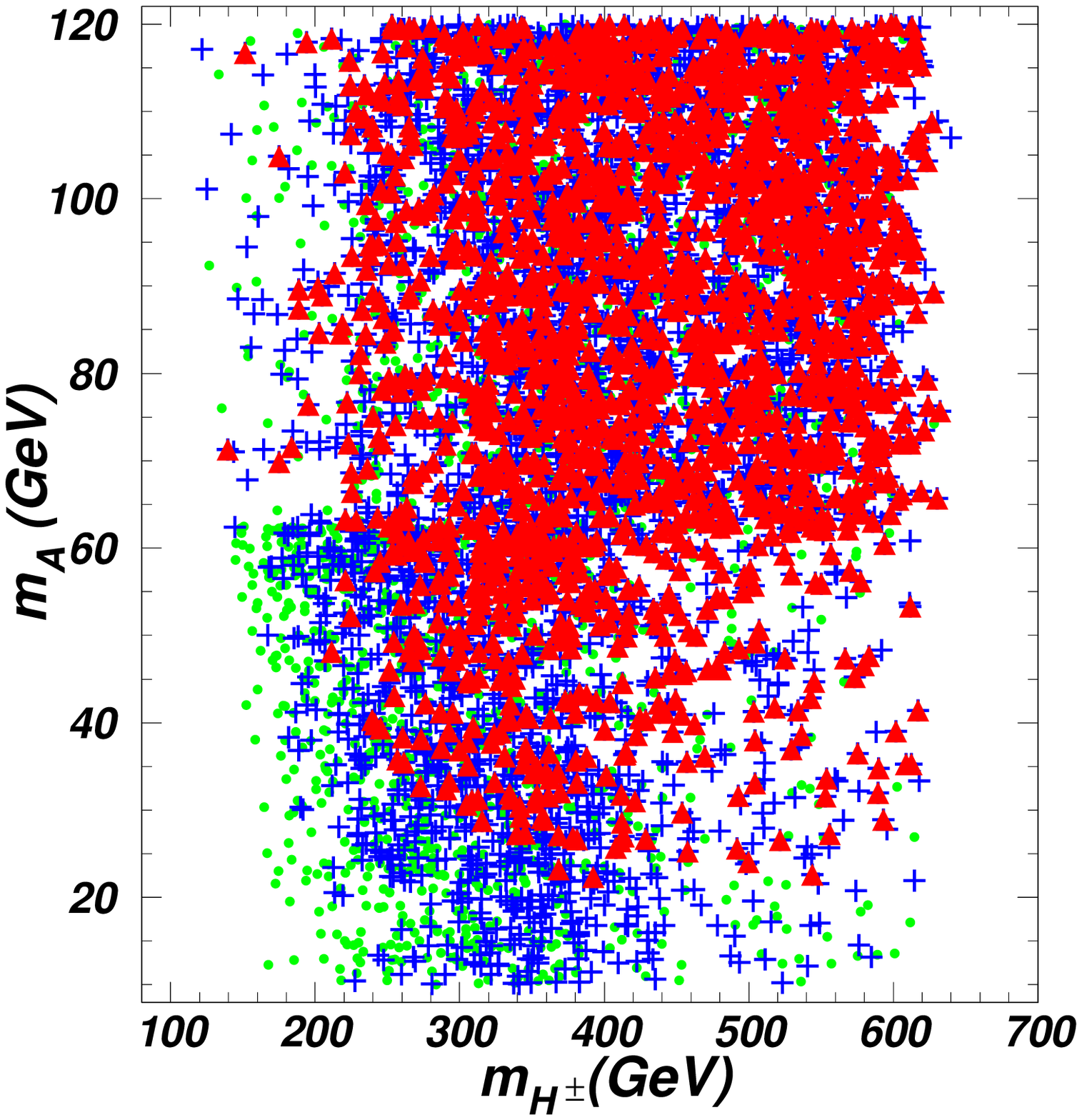,height=7.3cm}
  \epsfig{file=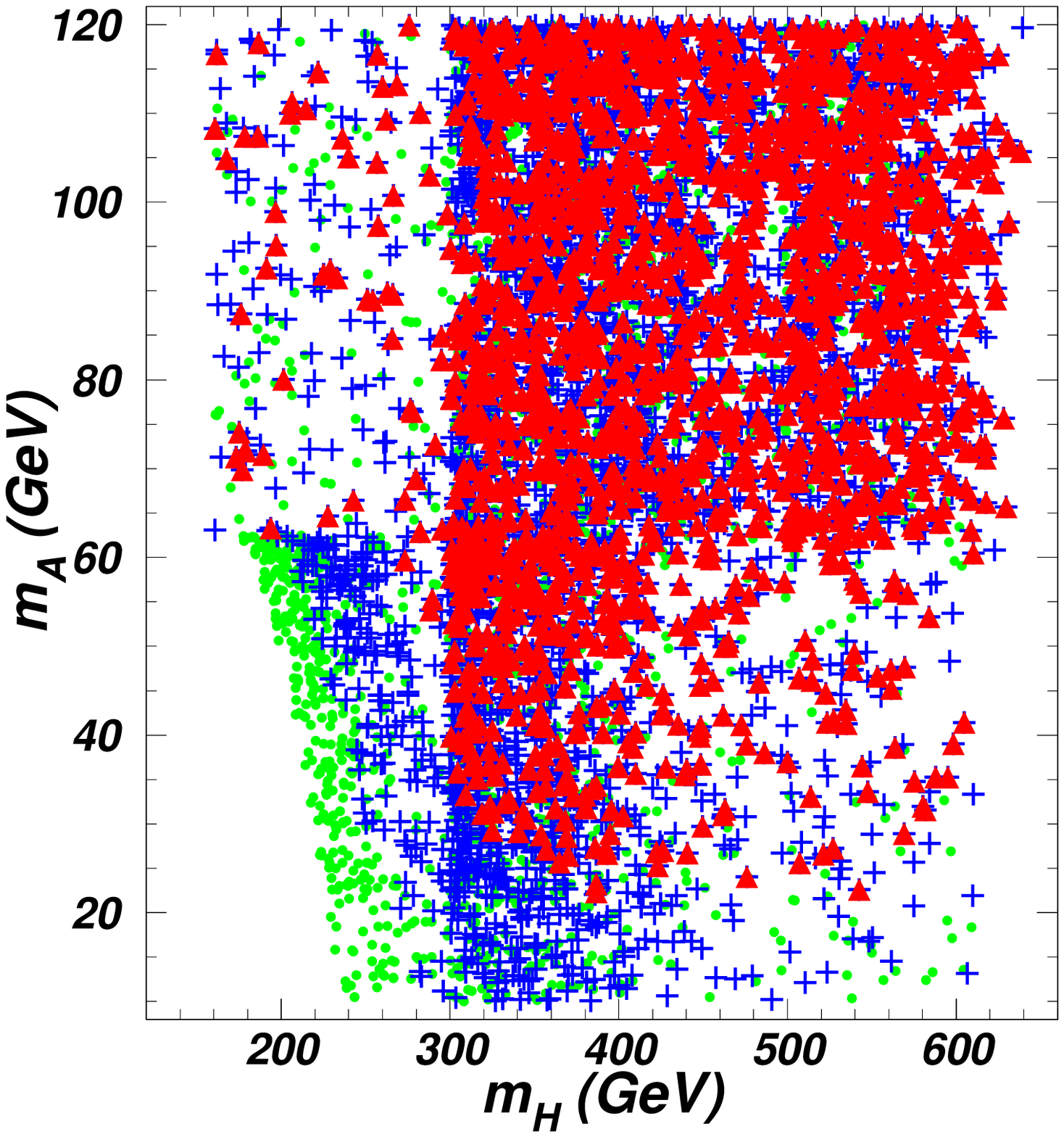,height=7.3cm}
\vspace{-0.5cm} \caption{The samples within the $1\sigma$, $2\sigma$
and $3\sigma$ ranges of $\Delta\chi^2$ projected on the
planes of $\tan\beta$ versus (VS) $m_{H^\pm}$, $\tan\beta$ VS $m_A$,  $m_A$ VS $m_{H^\pm}$, and
$m_A$ VS $m_H$ after imposing the constraints from theory, the oblique parameters, 
the exclusion limits from searches for Higgs at LEP,
the signal data of the 125 GeV Higgs, and the LFU from the $\tau$ decays.
The bullets (green), pluses (royal blue), and triangles (red) are respectively within 
the $3\sigma$, $2\sigma$ and $1\sigma$ regions of $\Delta\chi^2$.}
\label{chi123}
\end{figure}
%%%%%%%%%%%%%%%%%%%%

%%%%%%%%%%%%%%%%%%%%%
\begin{figure}[tb]
%\begin{center}
  \epsfig{file=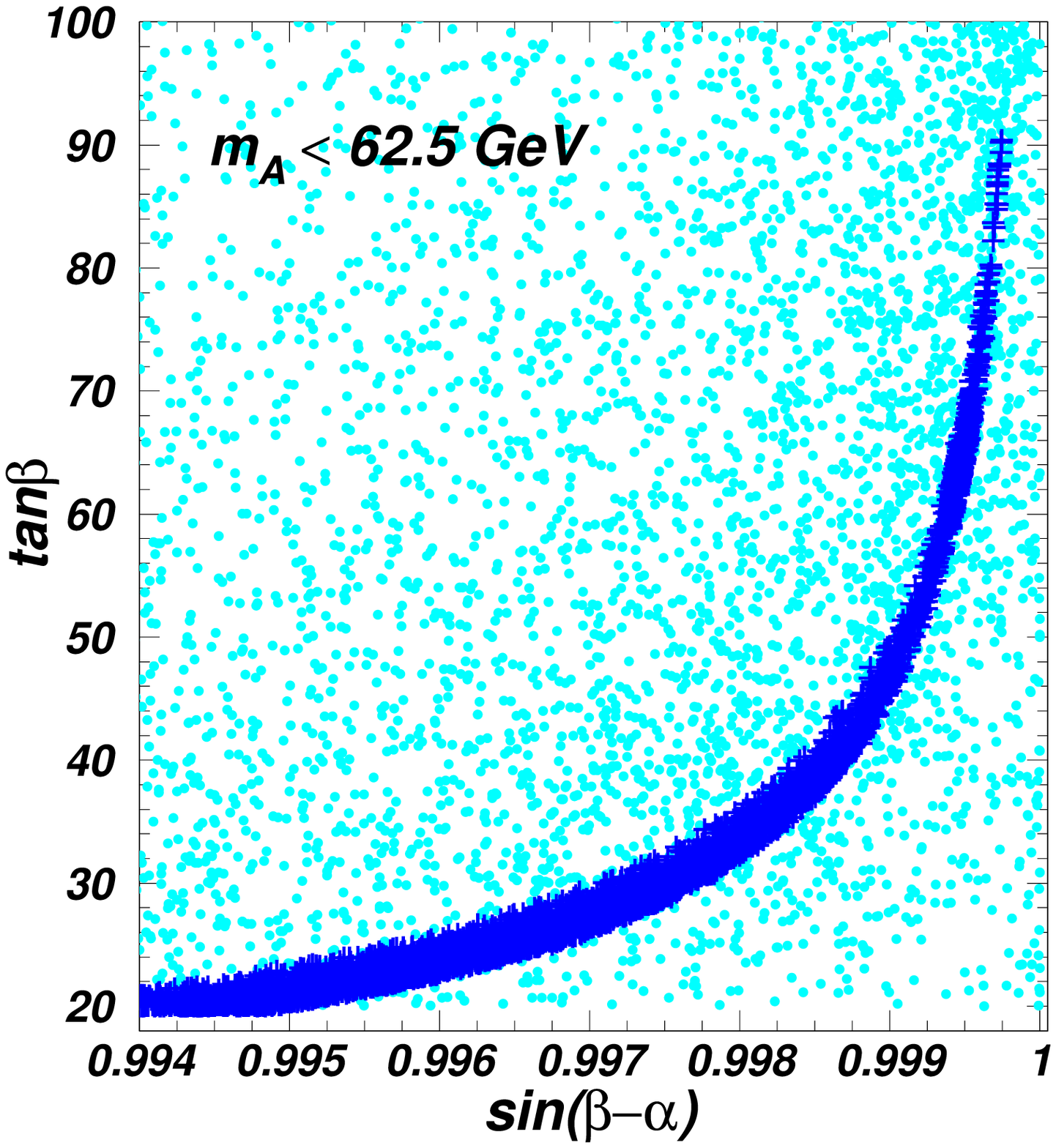,height=7.5cm}
  \epsfig{file=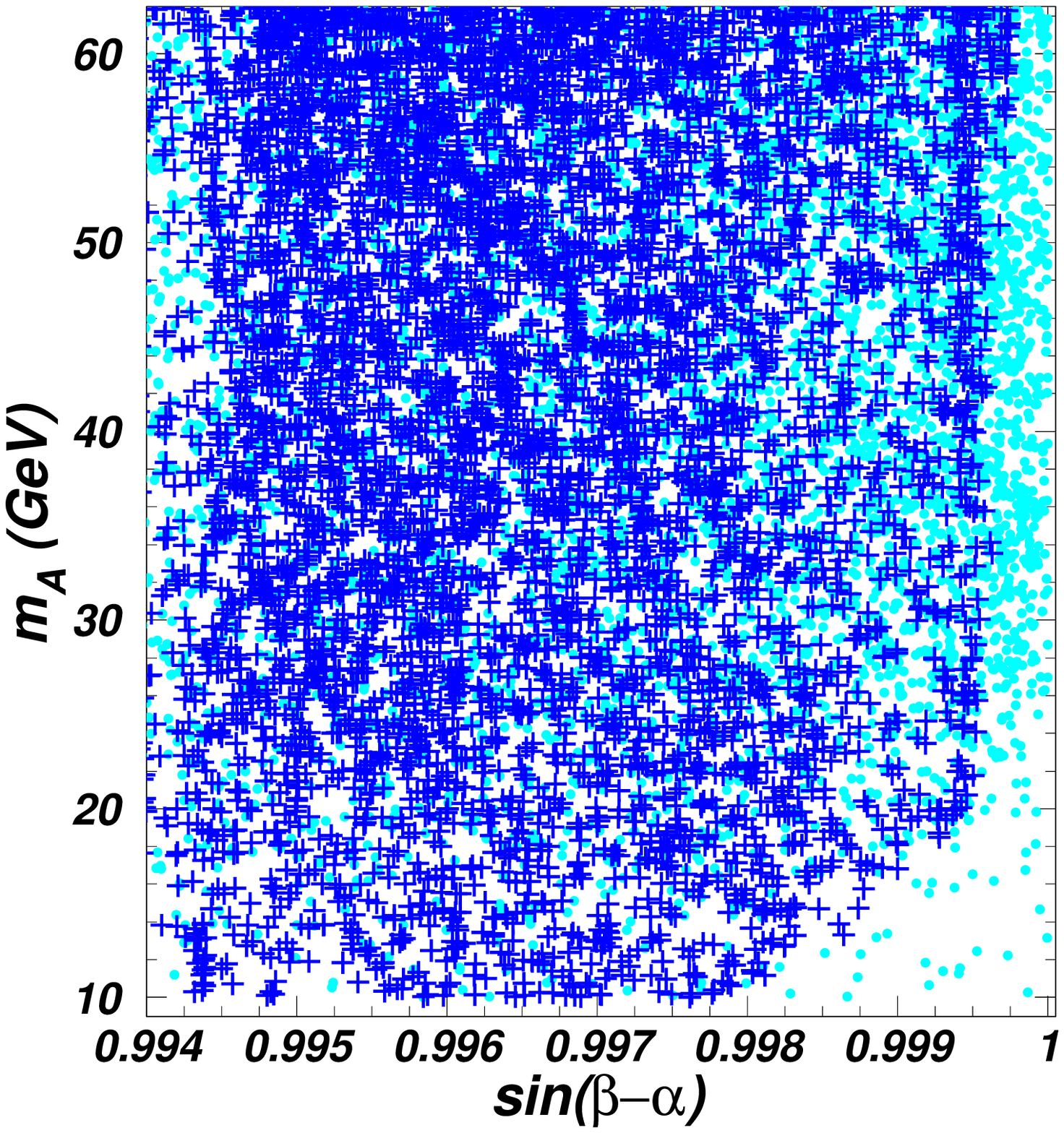,height=7.5cm}
\vspace{-0.5cm} \caption{The surviving samples projected on the planes
of $\sin(\beta-\alpha)$ VS $\tan\beta$ and $\sin(\beta-\alpha)$ VS $m_A$.
The bullets (sky blue) and pluses (royal blue) are allowed by the constraints from theory, 
the oblique parameters, the exclusion limits from searches for Higgses at LEP. In addition,
the pluses (royal blue) are within the $2\sigma$ regions of $\Delta\chi^2$.}
\label{sba}
\end{figure}
%%%%%%%%%%%%%%%%%%%%

%%%%%%%%%%%%%%%%%%%%%
\begin{figure}[tb]
%\begin{center}
  \epsfig{file=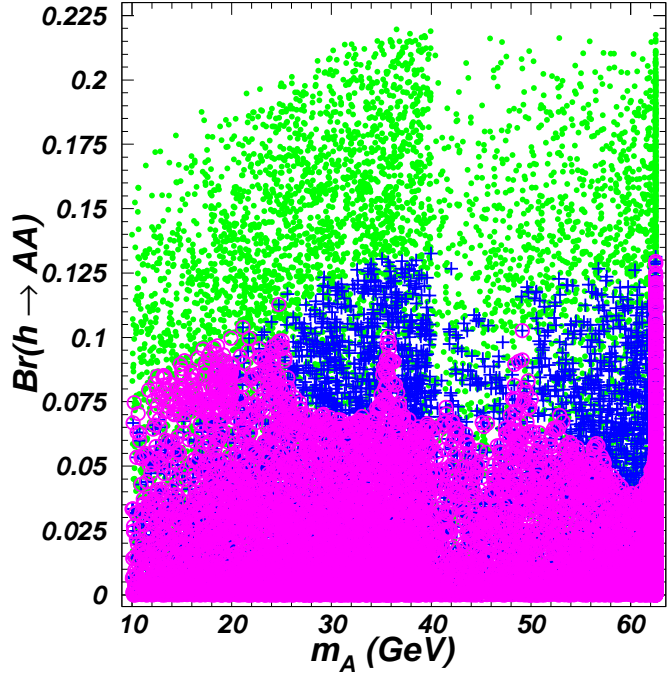,height=9cm}
\vspace{-0.5cm} \caption{The surviving samples on the planes
of Br$(h\to AA)$ VS $m_A$ after imposing the constraints from "pre-muon $g-2$". 
The bullets (green) and crosses (blue) 
 are respectively within the $3\sigma$ and $2\sigma$ regions of $\Delta\chi^2$. 
The circles (pink) are 
within the $2\sigma$ regions of $\Delta\chi^2$ and allowed by the exclusion limits 
from $h\to AA$ channels at the LHC.}
\label{haa}
\end{figure}
%%%%%%%%%%%%%%%%%%%%

%%%%%%%%%%%%%%%%%%%%%
\begin{figure}[tb]
%\begin{center}
  \epsfig{file=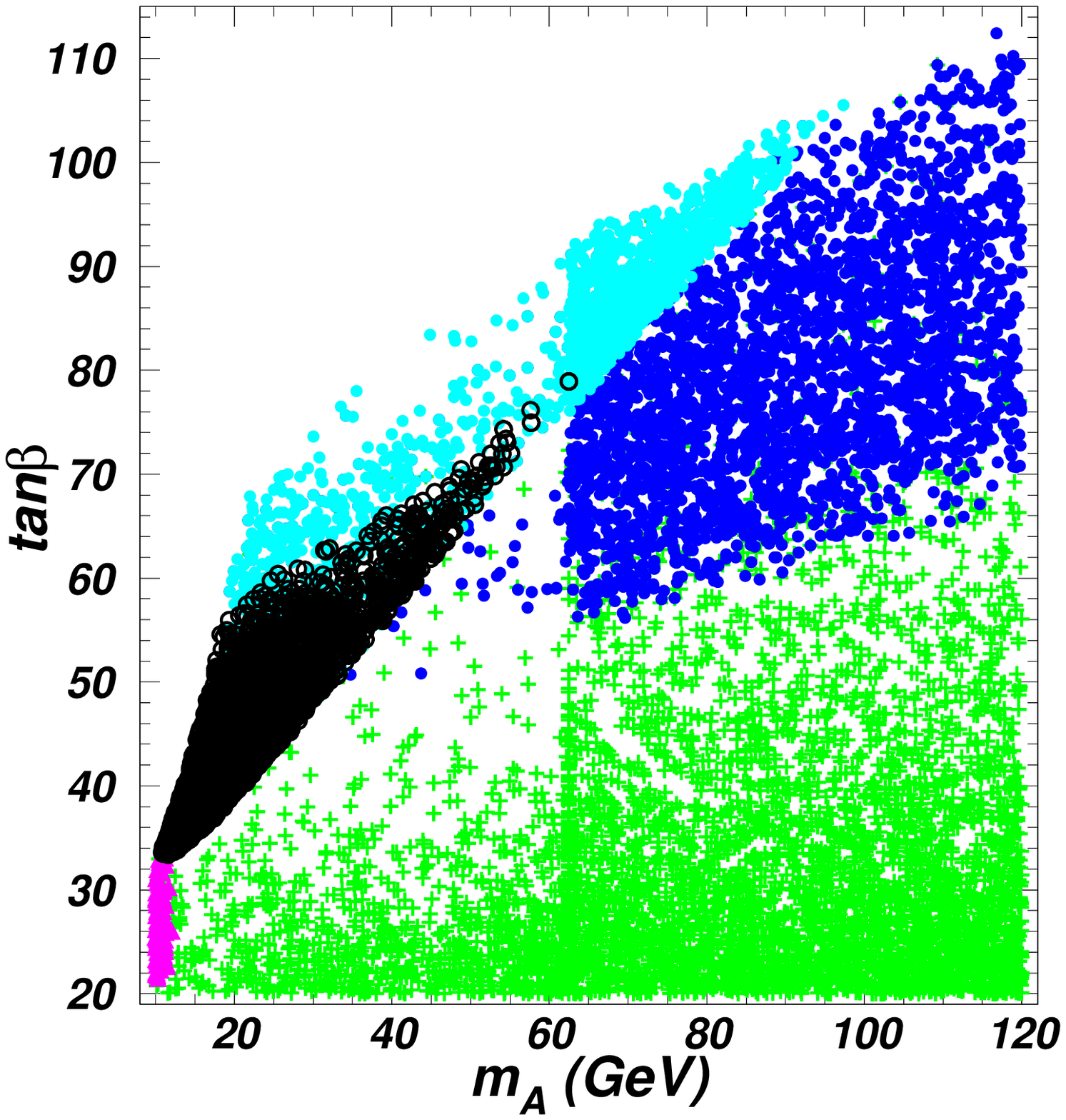,height=5.7cm}
  \epsfig{file=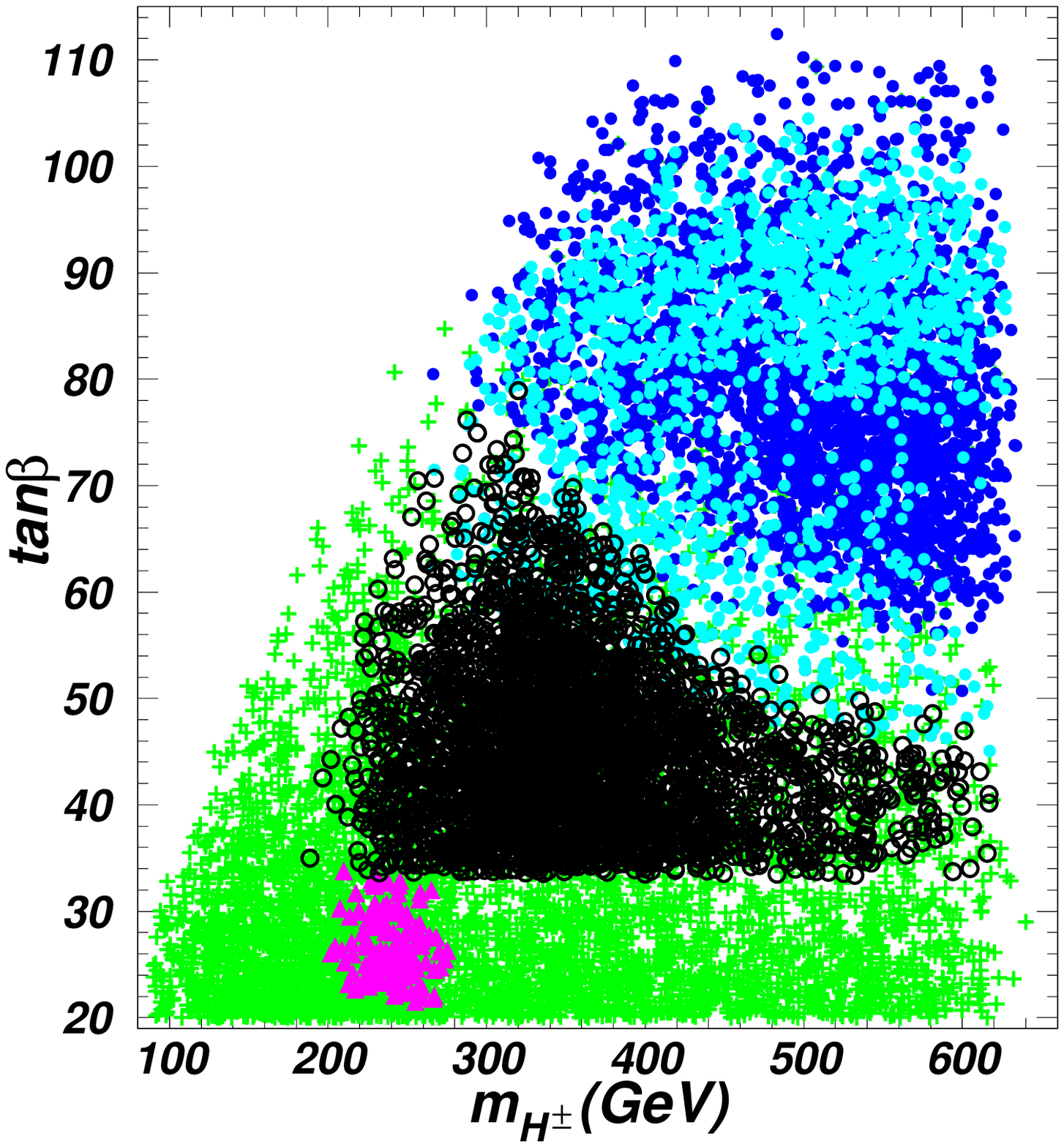,height=5.7cm}
  \epsfig{file=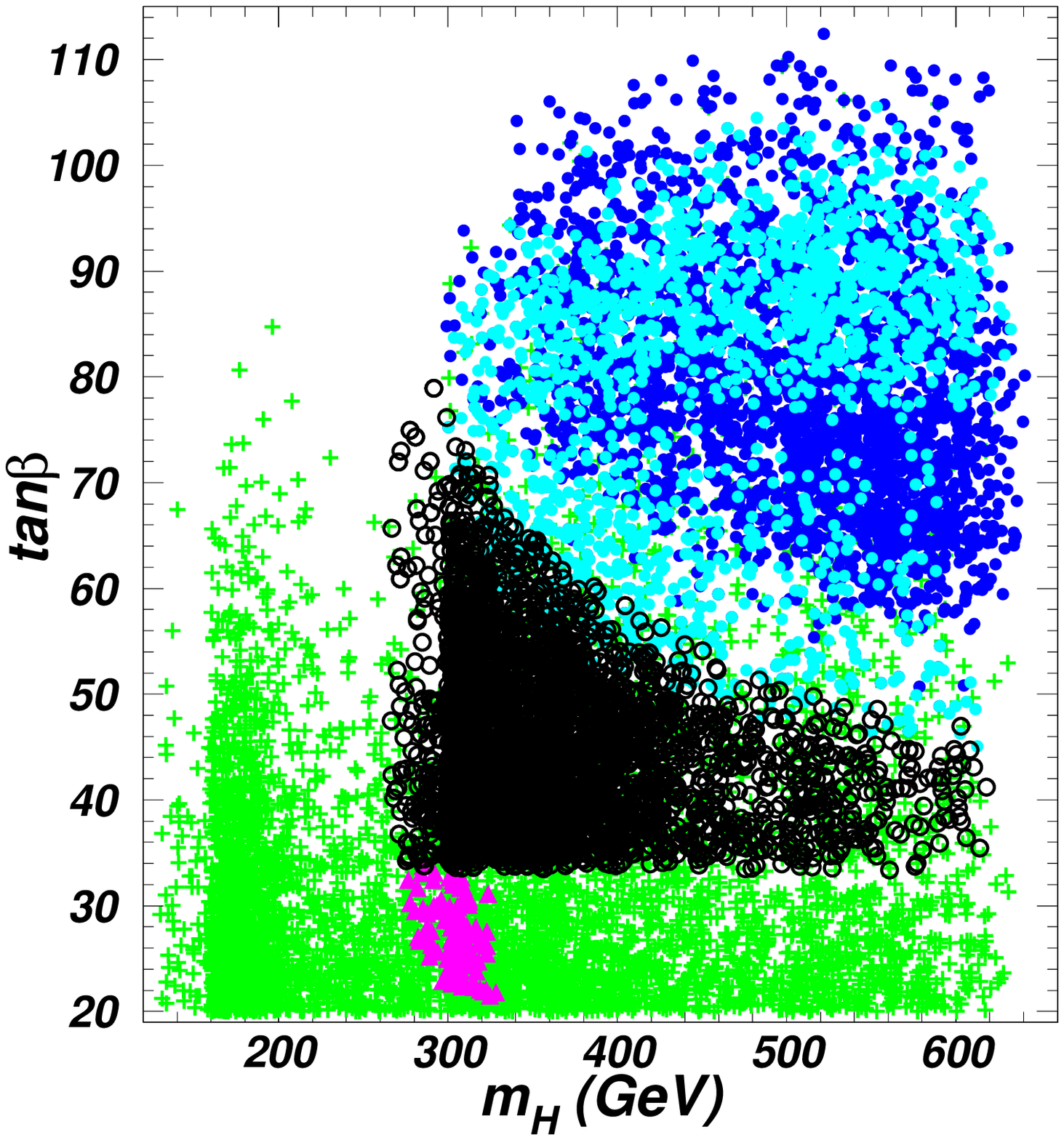,height=5.7cm}
  \epsfig{file=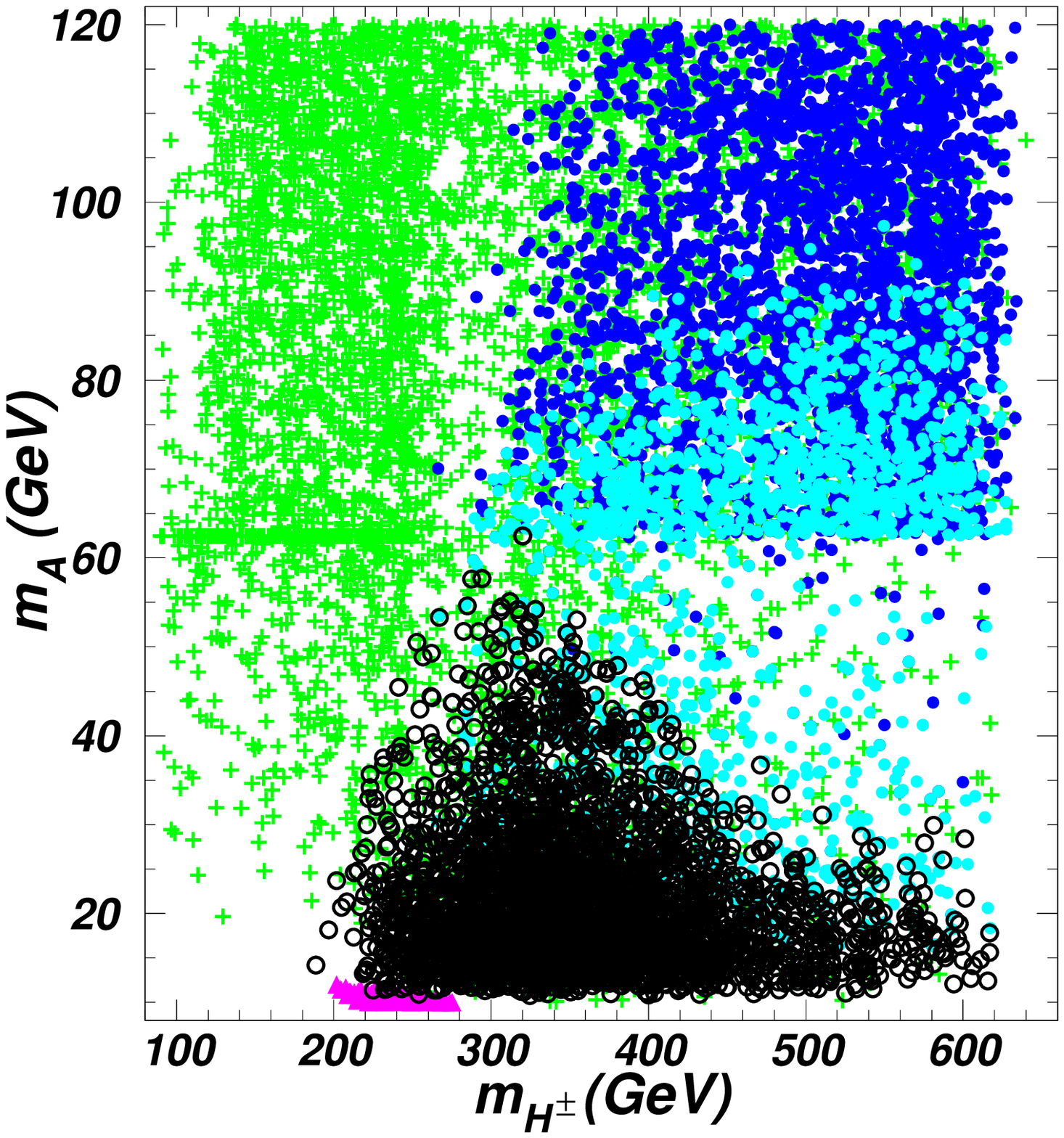,height=5.7cm}
  \epsfig{file=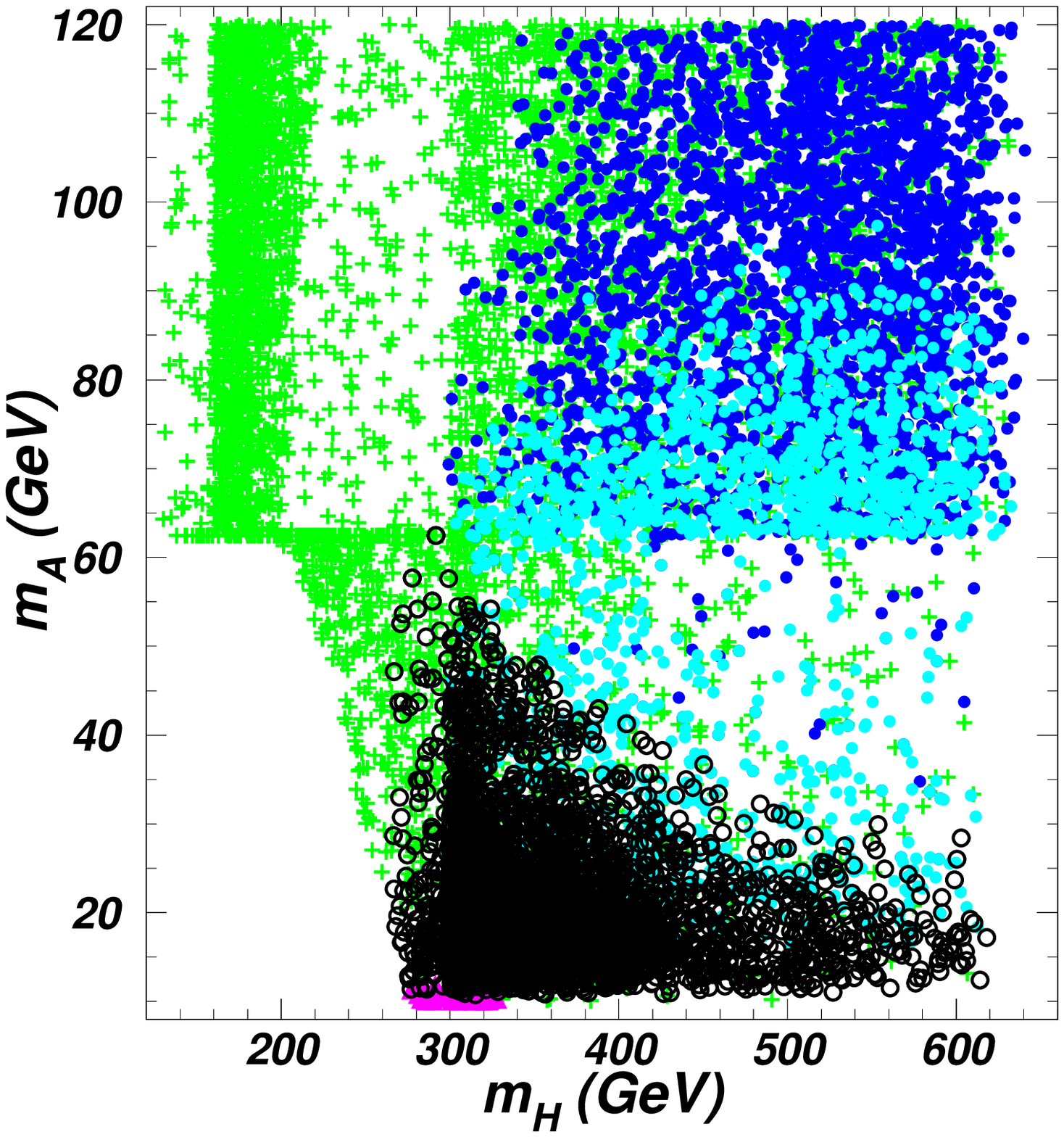,height=5.7cm}
  \epsfig{file=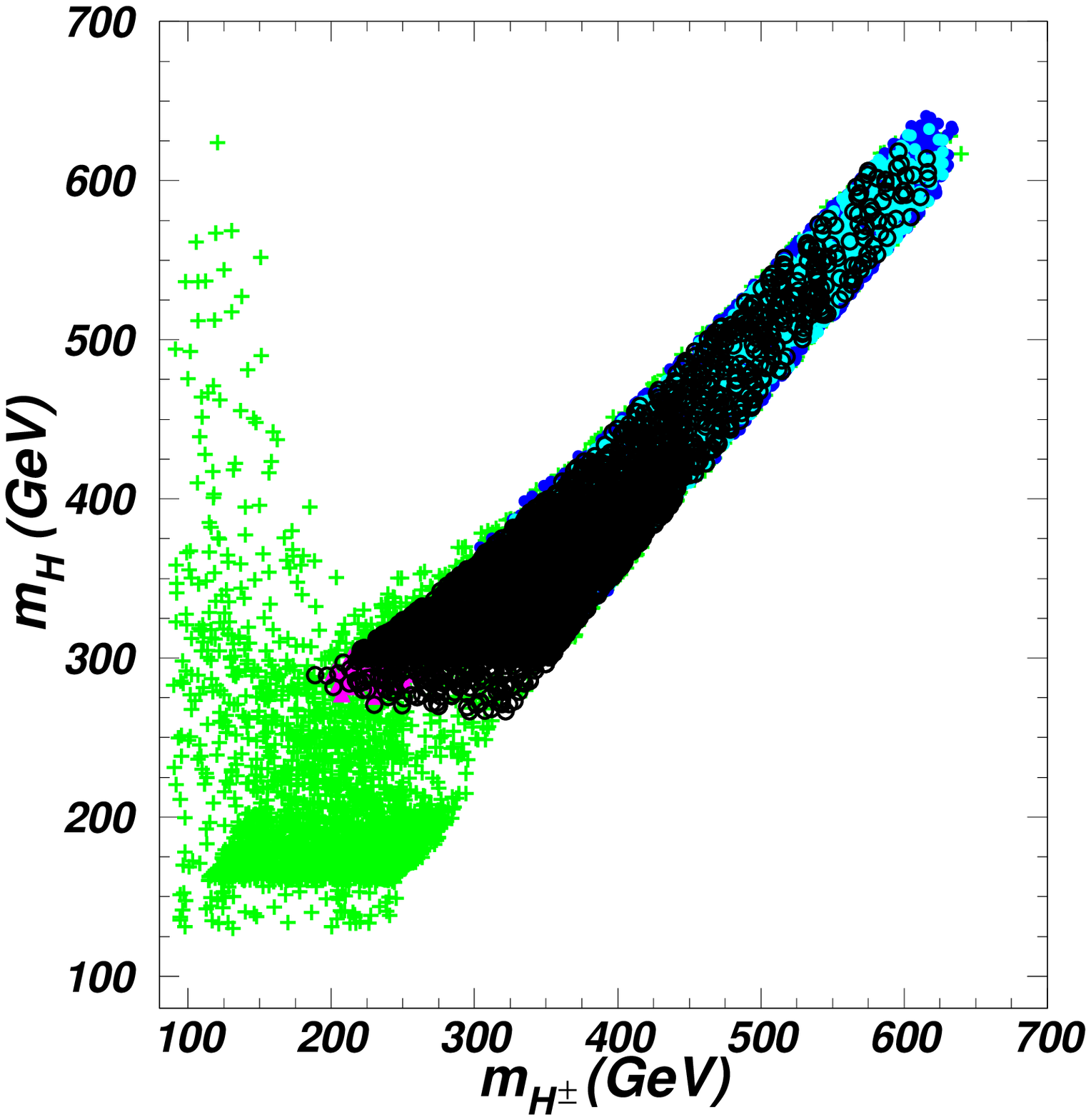,height=5.7cm}
\vspace{-0.25cm} \caption{The surviving samples projected on the
planes of $\tan\beta$ VS $m_A$, $\tan\beta$ VS $m_{H^\pm}$, 
$\tan\beta$ VS $m_H$, $m_A$ VS $m_{H^\pm}$,
$m_A$ VS $m_H$, and $m_H$ VS $m_{H^\pm}$. The pluses (green) are allowed by the constraints of "pre-muon $g-2$".
The triangles (pink) are allowed by the "pre-muon $g-2$", but excluded by the Br$(B_s\to \mu^+\mu^-)$ limits. 
The light bullets (sky blue) and dark bullets (royal blue) are allowed by the "pre-muon $g-2$", 
but excluded by the limits of the LFU in $Z$ decay.
In addition, the light bullets accommodate the muon $g-2$ anomaly and the dark bullets do not. The circles (black) are allowed by
the constraints from the "pre-muon $g-2$", the LFU in $Z$ decay, and Br$(B_s\to \mu^+\mu^-)$.}
\label{mug2}
\end{figure}
%%%%%%%%%%%%%%%%%%%

\subsection{Results and discussions}

In Fig. \ref{chi123}, we project the surviving samples 
within 1$\sigma$, 2$\sigma$, and 3$\sigma$ 
ranges of $\Delta\chi^2$ on the planes
of $\tan\beta$ VS $m_A$, $\tan\beta$ VS $m_{H^\pm}$, and $m_A$ VS
$m_{H^\pm}$ after imposing the constraints from theory, the oblique parameters, 
the exclusion limits from searches for Higgs at LEP,
the signal data of the 125 GeV Higgs, and the LFU in $\tau$ decays. 
We obtain a surviving sample with a minimal value of $\chi^2$ fit to the 125 GeV Higgs 
signal data
and the LFU data in $\tau$ decays, $\chi^2_{min}=16.99$. 
The upper-left panel of Fig. \ref{chi123} shows that the value of $\chi^2$ is favored to
increase with $\tan\beta$ and with a decrease of $m_{H^\pm}$. This is because 
the LFU in $\tau$ decays is significantly corrected by the tree-level diagrams mediated by the charged Higgs.
The lower-left panel of Fig. \ref{chi123} shows that $\chi^2$ is favored to 
have a large value for a small $m_A$. For $m_A< m_{H^\pm}$, the large mass splitting between 
$m_A$ and $m_{H^\pm}$
can make the one-loop diagram to give sizable correction to the LFU in 
$\tau$ decays.
The upper-right panel shows that, for a light pseudoscalar, such as $m_A<$ 25 GeV, $\tan\beta$ 
is strongly imposed an upper limit.
The main constraints are from the exclusion limits from the searches for Higgs at LEP. 
Ref. \cite{mu2h20} also obtained the limits from LEP on $\tan\beta$ for $m_A<$ 25 GeV. Our results
are consistent with those of Ref. \cite{mu2h20}, such as $\tan\beta>35$ (60) for $m_A$ around 10 GeV (20 GeV).
Most of regions of $m_A<$ 60 GeV and $m_{H}<$ 300 GeV are beyond the $2\sigma$ range 
of $\Delta\chi^2$, as shown in lower-right panel of Fig. \ref{chi123}. The main 
constraints are from the signal data of the 125 GeV Higgs and the theory.

The left panel of Fig. \ref{sba} shows that $\sin(\beta-\alpha)$ and $\tan\beta$ are restricted to be a narrow region.
From the Eq. (\ref{hffcoupling}), the $\tau$ Yukawa coupling of the 125 GeV Higgs normalized to the SM is
$y^h_\tau=\sin(\beta-\alpha)-\cos(\beta-\alpha)\tan\beta$ with $\cos(\beta-\alpha) > 0$, and $\mid y^h_\tau\mid$ can significantly 
deviate from 1.0 for a large $\tan\beta$, which is disfavored by the signal data of the 125 GeV Higgs. 
Therefore, an appropriate $\sin(\beta-\alpha)$ is required to make $\mid y^h_\tau\mid$ to be around 1.0.
A simple solution is $\mid\sin(\beta-\alpha)\mid$ very close to 1.0. However, the constraints of theory
and $Br(h\to AA)<0.3$ require $\tan\beta < 10$, and the detailed discussions are given in Ref. \cite{1412.3385} 
(See Fig. 1 of Ref. \cite{1412.3385}).
Therefore, such solution is excluded for $\tan\beta > 20$ in our paper, as shown in the right panel of Fig. \ref{sba}.
For $\sin(\beta-\alpha)>0$, $\mid\sin(\beta-\alpha)\mid$ is allowed to deviate from 1.0 properly. The corresponding
$-y^h_\tau$ is around 1.0, and $y^h_\tau$ is opposite in sign from the gauge boson coupling of the 125 GeV Higgs. This is so called 
the wrong sign Yukawa coupling of 125 GeV Higgs. For example, $y^h_\tau=-1.01$ for $\sin(\beta-\alpha)=0.999$ and $\tan\beta=45$.
For $\sin(\beta-\alpha)<0$, $y^h_\tau$ is smaller than 0, and its absolute value deviates from 1.0 significantly, which is excluded.
In addition, for $m_A<20$ GeV, the exclusion limits from the searches for Higgs at LEP impose an upper bound on $\tan\beta$.
As a result, some large values of $\sin(\beta-\alpha)$ are excluded for $m_A<20$ GeV according to the dependence of 
$\tan\beta$ on $\sin(\beta-\alpha)$ shown in the left panel.

In Fig. \ref{haa}, we project the surviving samples on the planes
of Br$(h\to AA)$ VS $m_A$
 after imposing the constraints from "pre-muon $g-2$" (denoting the theory, 
the oblique parameters, the exclusion limits from the searches for Higgs at LEP,
the signal data of the 125 GeV Higgs, the LFU in $\tau$ decays,
and the exclusion limits from $h\to AA$ channels at LHC). 
The direct searches for $h\to AA$ channels at the LHC
impose stringent upper limits on Br$(h\to AA)$ in the L2HDM, such 
as Br$(h\to AA)<4\%$ for $m_A=$ 60 GeV.
Many samples within the 2$\sigma$ range of $\Delta\chi^2$ are excluded.

In Fig. \ref{mug2}, we project the surviving samples on the planes
of $\tan\beta$ VS $m_A$, $\tan\beta$ VS $m_{H^\pm}$, $\tan\beta$ VS $m_{H}$,
$m_A$ VS $m_{H^\pm}$, $m_A$ VS $m_{H}$, and $m_H$ VS $m_{H^\pm}$ 
 after imposing the constraints from "pre-muon $g-2$", muon g-2 anomaly, 
the LFU in $Z$ decays, and
Br$(B_s\to \mu^+\mu^-)$. The lower-left and lower-middle panels show that
the LFU in $Z$ decays excludes most of samples in the 
large $m_{H^\pm}$ and $m_{H}$ regions. This is because that the one-loop diagram can give sizable corrections to the 
LFU in $Z$ decays for $m_A<m_{H^\pm}~(m_H)$.
The characteristic is also found in Refs. \cite{mu2h6,mu2h16,mu2h20}, and our results
are consistent with theirs.  
Because of the constraints from the oblique parameters, $H$ and $H^\pm$ are 
favored to have a small splitting mass
for large $m_{H^\pm}$, as shown in the lower-right panel of Fig. \ref{mug2}. 
Ref. \cite{mu2h8} also pointed out the constraints of $\Delta \rho$ on the 
the mass splitting between $H$ and $H^\pm$. The $T$ parameter used in our paper is related to $\Delta \rho$.
For $m_{H^\pm}<$ 250 GeV, all the samples within 2$\sigma$ region of $\Delta\chi^2$ are 
consistent with the 
limits of the LFU in $Z$ decay. This is because the LFU 
in $\tau$ decays
can give more stringent constraints on the L2HDM than the LFU in $Z$ 
decays for a light charged Higgs. 

The upper-left and lower-left panels of Fig. \ref{mug2} show that the limits of Br$(B_s\to \mu^+\mu^-)$ 
exclude most of regions of $m_A$ around 10 GeV and $m_{H^\pm}<$ 300 GeV.
The $A$ exchange diagrams can give sizable contributions to $B_s\to \mu^+\mu^-$ for a very 
small $m_A$.
In the L2HDM, the lepton couplings are enhanced by $\tan\beta$, while the quark couplings
are suppressed by $\cot\beta$. Therefore, the leading contributions are almost independent on
$\tan\beta$ for large $\tan\beta$. 

Fig. \ref{mug2} shows that with the limits from "pre-muon $g-2$", the LFU 
in $Z$ decays and Br$(B_s\to \mu^+\mu^-)$ being satisfied, 
the muon $g-2$ anomaly can be explained in the regions of 32 $<\tan\beta<$ 80, 
10 GeV $<m_A<$ 65 GeV,
260 GeV $<m_H<$ 620 GeV, and 180 GeV $<m_{H^\pm}<$ 620 GeV. 
The upper-left panel of Fig. \ref{mug2} shows that
in the range of 65 GeV $<m_A<$ 100 GeV, the muon $g-2$ anomaly can be explained 
for a large enough $\tan\beta$.
However, such a large $\tan\beta$ is excluded by the LFU in $Z$ decays. 
 The contributions of $A$ to the muon $g-2$ anomaly have destructive 
interference with those of $H$. Therefore, a large mass splitting between $A$ and $H$ 
is required to explain the muon $g-2$ anomaly,
as shown in the lower-middle panel of Fig. \ref{mug2}.

\section{The direct search limits from the LHC}
Here we discuss the direct search limits from the LHC.
In the parameter space in favor of muon $g-2$ anomaly explanation, 
the production processes of extra Higgs bosons
via the Yukawa interaction with quarks can be neglected due to the suppression 
of large $\tan\beta$ in the L2HDM.
For $m_A$ smaller than 62.5 GeV, a pair of pseudoscalars can be produced 
via $pp\to h \to AA$ at the LHC.
In the above Section, we find that $h\to AA$ channel at the LHC can exclude many samples
within the 2$\sigma$ region of $\Delta\chi^2$.

%%%%%%%%%%%%%%%%%%%%%
\begin{figure}[thb]
%\begin{center}
  \epsfig{file=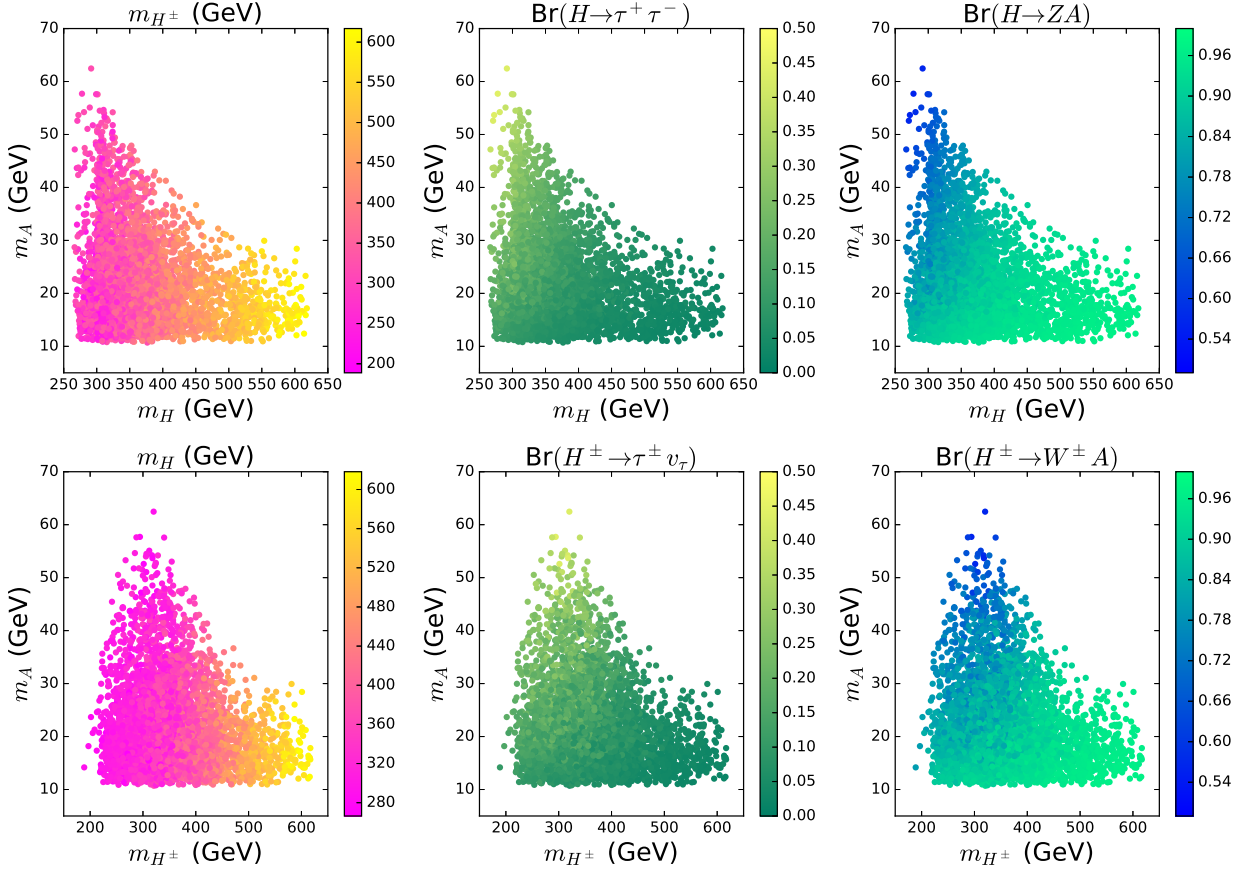,height=12cm}
\vspace{-1.5cm} 
\caption{The samples satisfying the constraints described in Sec.~\ref{constraints} 
projected on the planes of $m_{H}$ VS $m_{A}$ and $m_{H^\pm}$ VS $m_{A}$. 
The varying colors in each panel indicate the values of $m_{H^\pm}$, BR$(H\to\tau^{+}\tau^{-})$,
BR$(H\to ZA)$, $m_{H}$, BR$(H^\pm\to\tau^{\pm}v_{\tau})$ and BR$(H^\pm\to W^{\pm}A)$, respectively.}
\label{lhc1}
\end{figure}
%%%%%%%%%%%%%%%%%%%%

The extra Higgs bosons are dominantly produced at the LHC via
the following electroweak processes:
\begin{align}
pp\to & W^{\pm *} \to H^\pm A, \label{process1}\\
pp\to &       Z^*/\gamma^* \to HA, \label{process2}\\
pp\to & W^{\pm *} \to H^\pm H, \label{process3}\\
pp\to & Z^*/\gamma^* \to H^+H^-. \label{process4}
\end{align}
In our scenario, the important decay modes of the Higgs bosons are 
\begin{align}
A\to  \tau^+\tau^-,~\mu^+\mu^-,~~ H\to  \tau^+\tau^-,~ZA,~~
H^\pm\to  \tau^\pm\nu,~W^\pm A.
\end{align}
Here the light pseudo-scalar $A$ indeed decays into $\tau \tau$ essentially at 100\% due to the enhanced lepton Yukawa couplings by large $\tan\beta$. The other decay branch ratios and mass spectrum for the samples satisfying constraints described in Sec.~\ref{constraints} are presented in Fig.~\ref{lhc1} on the planes of $m_{H}$ VS $m_{A}$ and $m_{H^\pm}$ VS $m_{A}$. We can see from the upper panels that $m_{H}$ increases from 260 GeV to 620 GeV with $m_{H^\pm}$ increasing from 180 GeV to 620 GeV and the upper bounds of $m_A$ decreasing from 65 GeV to 30 GeV. The reason are discussed in Sec.~\ref{constraints}. As a result, the cross sections of processes in Eq.~(\ref{process1})  and Eq.~(\ref{process2}) are much larger than the two in Eq.~(\ref{process3}) and Eq.~(\ref{process4}). 
 The middle and right panels exhibit the decay branch ratios of $H/H^\pm$ to $\tau^+\tau^-/\tau^\pm v_{\tau}$ and $H/H^\pm$ to gauge boson and $A$. With an increase of $m_A$, the partial widths of $H^\pm/H$ to $A W^\pm/Z$ 
decrease due to the suppression of phase space. The muon $g-2$ anomaly favors a large $\tan\beta$ with $m_A$ increasing, which leads
the partial widths of $H\to\tau^{+}\tau^{-}$ and $H^\pm\to\tau^{\pm}v_{\tau}$ to be enhanced since the Yukawa couplings are 
proportional to $\tan\beta$. Therefore, with an increase of $m_A$, Br$(H\to AZ)$ and Br$(H^\pm\to W^\pm A)$ decrease, 
and Br$(H\to\tau^{+}\tau^{-})$ and Br$(H^\pm\to\tau^{\pm}v_{\tau})$ increase.
 In conclusion, the dominated finial states generated at LHC of our samples are 3 or 4 $\tau$s with or without gauge boson from
\begin{align}
pp\to & W^{\pm *} \to H^\pm A \to , 3\tau+v_{\tau} ~{\rm or }~ 4\tau+W^\pm \label{finalstate1}\\
pp\to &       Z^*/\gamma^* \to HA \to 4\tau ~{\rm or }~ 4\tau+Z \label{finalstate2}.
\end{align}
In order to restrict the productions of the above processes at the LHC for our model, 
we perform simulations for the samples in Fig.~\ref{lhc1} 
using \texttt{MG5\_aMC-2.4.3}~\cite{Alwall:2014hca} 
with \texttt{PYTHIA6}~\cite{Torrielli:2010aw} and 
\texttt{Delphes-3.2.0}~\cite{deFavereau:2013fsa}, and adopt the constraints 
from all the analysis for the 13 TeV LHC in 
version \texttt{CheckMATE 2.0.7}~\cite{Dercks:2016npn}. 
Besides, the latest multi-lepton searches for 
electroweakino~\cite{Sirunyan:2017zss,Sirunyan:2017qaj,Sirunyan:2017lae,Sirunyan:2018ubx,Aaboud:2017nhr} implemented in Ref.~\cite{Pozzo:2018anw} are also taken into consideration 
because of the dominated multi-$\tau$ final states in our model.

%%%%%%%%%%%%%%%%%%%%%
\begin{figure}[thb]
%\begin{center}
  \epsfig{file=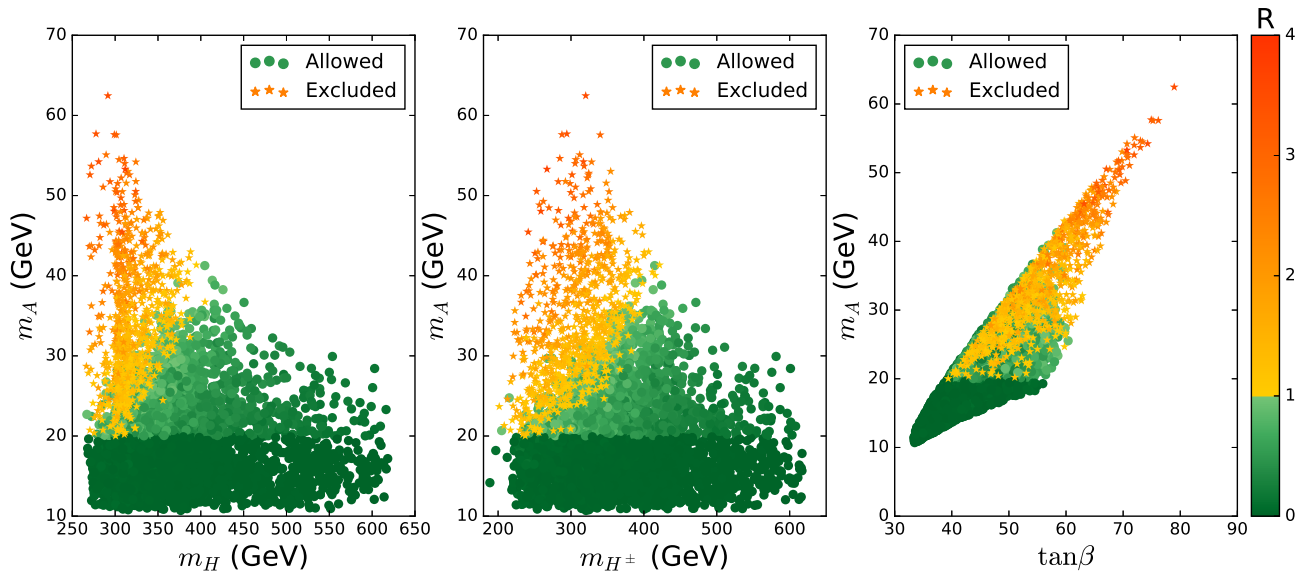,height=8cm}
\vspace{-1.0cm} 
\caption{The samples satisfying the constraints described in Sec.~\ref{constraints}, 
projected on the planes of 
$m_{H}$ VS $m_{A}$, $m_{H^\pm}$ VS $m_{A}$, and $\tan\beta$ VS $m_A$ with colors 
indicating the \texttt{R} values from \texttt{CheckMATE}. The orange stars and green dots 
stand for the samples excluded and allowed by the LHC Run-2 data at 95\% confidence level, 
respectively.}
\label{lhc2}
\end{figure}
%%%%%%%%%%%%%%%%%%%%

The results from \texttt{CheckMATE} are presented in Fig.~\ref{lhc2} on the 
planes of $m_{H}$ VS $m_{A}$, $m_{H^\pm}$ VS $m_{A}$,
 and $\tan\beta$ VS $m_A$. The colors stand for the \texttt{R} values 
defined by \cite{Dercks:2016npn}
\begin{align}
\texttt{R}=\max_i \{ \frac{S_i-1.96\Delta S_i}{S_{i,{\rm Exp}}^{95}}\},
\end{align}
where $S_i$ and $\Delta S_i$ denote the number of signal events in signal region $i$, and $S_{i,{\rm Exp}}^{95}$ is the experimentally measured 95\% confidence limit on signal event in signal region $i$. Obviously, \texttt{R} $>1$ means that the corresponding point is excluded at 95\% confidence level by at least one search channel. We can see that the constraints from current LHC 13 TeV data shrink $m_A$ from [10, 65] GeV to [10, 44] GeV and $\tan\beta$ from [32, 80] to [32, 60]. For the samples excluded by current 13 TeV LHC data, the strongest constraint comes from the search for electroweak production of charginos and neutralinos in multilepton final states \cite{Sirunyan:2017lae}. In this analysis, 7 categories of signal region are designed for event with $\tau$ in final state, \texttt{SR-C} to \texttt{SR-F} and \texttt{SR-I} to \texttt{SR-K}. The most sensitive signal region is \texttt{SR-K} for most of the parameter space. It requires at least two light-flavor leptons and two $\tau$ jets without b-tagged jet. The signal region is subdivided by missing energy $E\!\!\!/_T$ to three bins, \texttt{SR-K01}, \texttt{SR-K02}, and \texttt{SR-K03}. The main contributions of our samples to the bins are from processes in Eq.~(\ref{finalstate1}) and Eq.~(\ref{finalstate2}) with at least two of the $\tau$s decaying hadronically.

In Fig.~\ref{lhc2}, the points with relatively larger $m_{H}$/$m_{H^\pm}$ or lower $m_{A}$ can escape the direct searches. The \texttt{R} value decreases gently with heavier $H$/$H^\pm$ because of the smaller cross sections. With higher luminosity and collision energy this region can be further detected. For the light $A$, the $\tau$s from $A$ in
Eq.~(\ref{process1}) to Eq.~(\ref{process4}) decays become too soft to be distinguished at detector, while the $\tau$s from $A$ in $H/H^\pm$ decays are collinear because of the large mass splitting between $A$ and $H/H^\pm$. Meanwhile, the $H/H^\pm \to A Z/W^\pm$ decay modes dominate the $H/H^\pm$ decays in the low $m_{A}$ region. Thus, in the region of $m_{A}<20$ GeV, the acceptance of above signal region for final state of two collinear $\tau$ + $Z/W$ boson quickly decreases.

The production processes of the extra Higgses in Eqs. (\ref{finalstate1}, \ref{finalstate2}) 
considered by us are the same as Eqs. (26-29) of
  Ref. \cite{mu2h11}. The main difference is that we implemented the constraints from 13 TeV LHC 
results of 36 fb$^{-1}$ data, while Ref. \cite{mu2h11} used the 8 TeV LHC results of  20 fb$^{-1}$ data. 
Another difference is that we perform MC simulation for all survived samples instead of 
points on $m_A-m_H$ plane with fixed $m_{H^{\pm}}$ and branch ratios. In addition, the constraints of LFU
 in $Z$ decay are not considered in Ref. \cite{mu2h11}. Thus, $m_H$ and $m_{H^\pm}$ are allowed to be 
large enough to satisfy the LHC searches for $HA$ and $H{^\pm}A$ productions in Ref. \cite{mu2h11}.
 However, the lower panels of Fig. \ref{mug2} in our paper
show that the limits of LFU in $Z$ decays can impose the upper bounds of $m_H$ and $m_{H^\pm}$ in
the parameter space favored by the muon $g-2$ anomaly, such as $m_H<320$ GeV for $m_A=50$ GeV. Such ranges of 
$m_H$ and $m_A$ are completely excluded by the LHC searches for $HA$ production, and the corresponding $\tan\beta$
is also excluded.

\section{The strong first-order phase transition}
In this section we study the possibility to obtain a parameter space in L2HDM 
that can trigger a SFOPT and explain muon $g-2$ anomaly at the same time. 
In order to know the strength of phase transition in our scenario, we need to study the effective potential with thermal correction included.  
The thermal effective potential $V(\phi_1,\phi_2,T)$ at temperature $T$ is composed of four parts:
\begin{eqnarray}
V(\phi_1,\phi_2,T) = V^{0}(\phi_1,\phi_2) + V^{CW}(\phi_1,\phi_2) + V^{CT}(\phi_1,\phi_2) + V^{T}(\phi_1,\phi_2,T) ,
\end{eqnarray}
where $V^{0}$ is the tree-level potential, $V^{CW}$ is the Coleman-Weinberg potential, 
$V^{CT}$ is the counter term and $V^{T}$ is the thermal correction. 
Concrete expressions of these terms can be found in~\cite{TEF}.

The condition for a SFOPT is usually taken to be~\cite{washout}: 
\begin{eqnarray}
\xi_c \equiv \frac{v_c}{T_c} \geqslant 1.0. 
\end{eqnarray}
Here $T_c$ is the critical temperature at which a second minimum of $V(\phi_1,\phi_2,T)$ with non-zero VEV appear, and $v_c = \sqrt{\phi_1^2+\phi_2^2}$ is the corresponding VEV at $T_c$.
Due to the complicated form of $V(\phi_1,\phi_2,T)$, numerical calculation is always required to analyze the geometry evolution of $V(\phi_1,\phi_2,T)$. 
In this work we use package \textsf{BSMPT}~\cite{Basler:2018cwe} to do the analysis. 
In \textsf{BSMPT}, the critical temperature $T_c$ is determined when the minimization point $v = v_c$ at critical temperature $T_c$ jumps to the origin $v = 0$ at a slightly higher temperature $T>T_c$.

%%%%%%%%%%%%%%%%%%%%%
\begin{figure}[thb]
%\begin{center}
  \epsfig{file=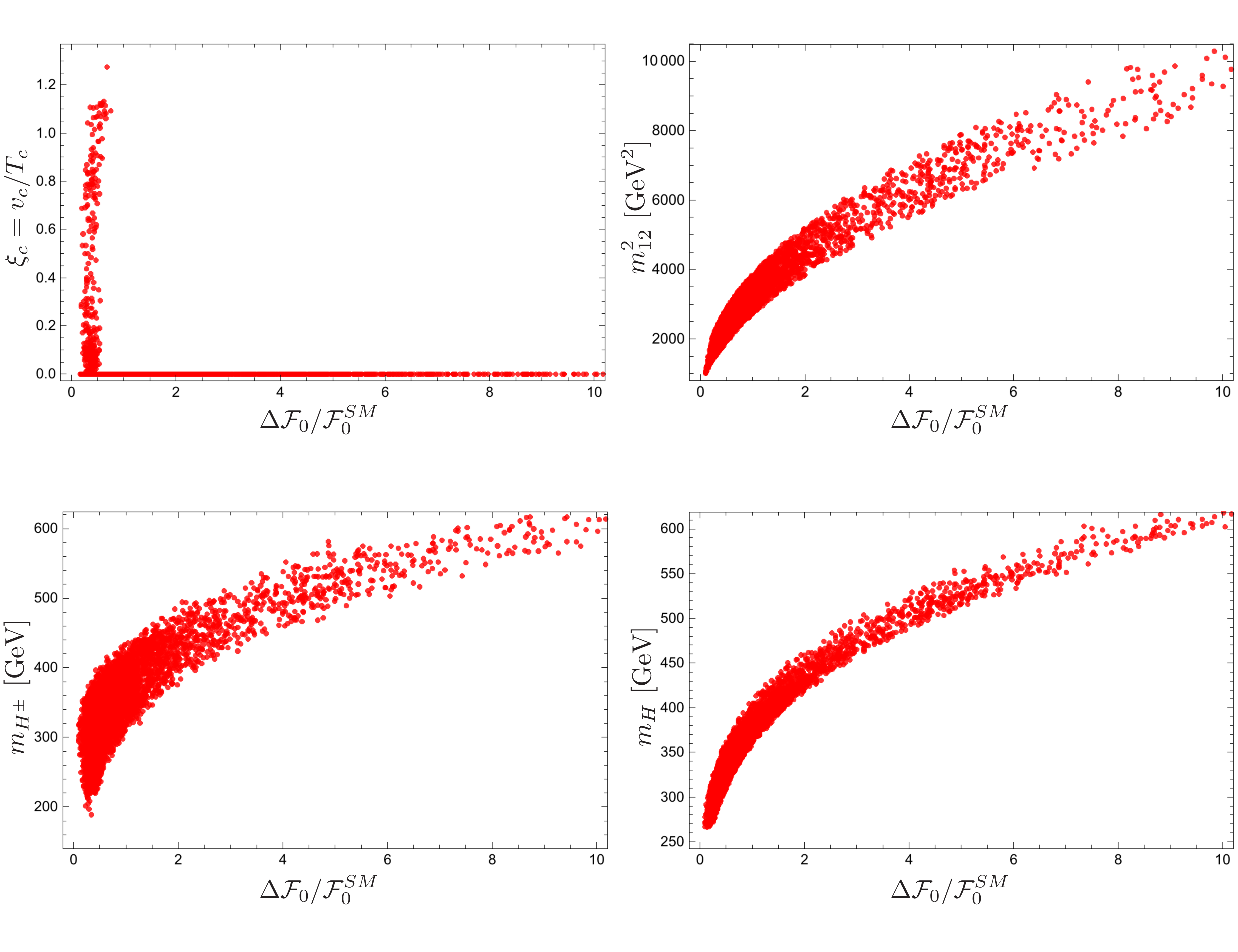,height=10cm}
\vspace{-1.0cm} 
 \caption{Upper-left: relationship between $\Delta \mathcal{F}_0 / \mathcal{F}^{SM}_0$ and $\xi_c$. 
Points with $\xi_c=0$ do not have a first order phase transition. Relationships between $\Delta \mathcal{F}_0 / \mathcal{F}^{SM}_0$ and $m^2_{12}$ (upper-right),
 $\Delta \mathcal{F}_0 / \mathcal{F}^{SM}_0$ and $m_{H^\pm}$ (lower-left), and $\Delta \mathcal{F}_0 / \mathcal{F}^{SM}_0$ and $m_H$ (lower-right). }
\label{SFOPT_1}
\end{figure}
%%%%%%%%%%%%%%%%%%%%

%%%%%%%%%%%%%%%%%%%%%
\begin{figure}[thb]
%\begin{center}
  \epsfig{file=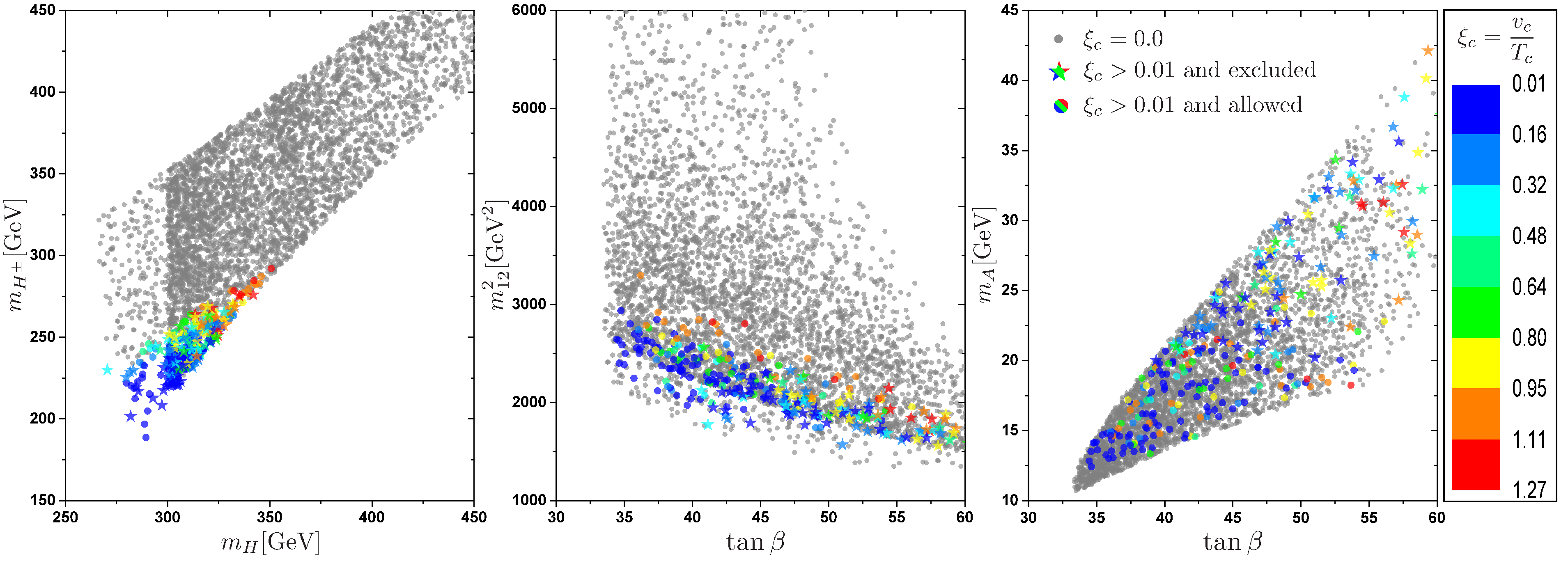,height=6cm}
\vspace{-1.0cm} 
\caption{$m_H$ VS $m_{H^\pm}$ plane (left panel), $\tan\beta$ VS $m^2_{12}$ (middle) and 
$\tan\beta$ VS $m_A$ (right) with color mapped by $\xi_c$. Grey points do not have a SFOPT. 
Colored spots are allowed points under current limits, and colored stars are excluded by the LHC direct search.}
\label{SFOPT_2}
\end{figure}
%%%%%%%%%%%%%%%%%%%%

The 4881 points allowed and excluded by the search limits of LHC in the Sec. IV are used as input to calculate $\xi_c$. 
Out of 4881 input points, there are only 279 points that can lead to VEV jumping and a non-zero $\xi_c$.
Because of the complicated scalar potential geometry and its dependence on $T$, it is hard to find obvious relation between our zero temperature inputs and the strength of phase transition. 
While in~\cite{Vacuum_PT,CPV}, it is pointed out that the depth of minimum point at zero temperature has a strong correlation with phase transition strength. 
If the zero temperature vacuum energy in a model (noted as $\mathcal{F}_0$) is higher than the zero temperature vacuum energy in the SM (noted as $\mathcal{F}^{SM}_0$), then the phase transition of this model tends to be SFOPT. A $\mathcal{F}_0$ under $\mathcal{F}^{SM}_0$ can also trigger a first order phase transition, but with a lower probability and a lower phase transition strength. 
The difference between $\mathcal{F}_0$ and $\mathcal{F}^{SM}_0$ at one-loop level in 2HDM with alignment limit ($\sin(\beta - \alpha) \to 1$) has been given in~\cite{CPV}:
\begin{eqnarray}
\nonumber \Delta \mathcal{F}_0 &\equiv&  \mathcal{F}_0 - \mathcal{F}^{SM}_0 = \frac{1}{64\pi^2} \left[ (m^2_h - 2 M^2)^2 \left( \frac{3}{2} + \frac{1}{2}\log \left[ \frac{4m_Am_Hm_{H^\pm}}{(m^2_h - 2 M^2)^2} \right]    \right) \right] \\
& & + \frac{1}{2}(m^4_A+m^4_H+2m^4_{H^\pm}) + (m^2_h - 2 M^2)(m^2_A+m^2_H+2m^2_{H^\pm})  \ .
\end{eqnarray}
Here $M^2 = m^2_{12} (\tan\beta + \tan\beta^{-1}) $.
The SM one-loop vacuum energy $\mathcal{F}^{SM}_0 \approx -1.25 \times 10^8 $GeV$^4$.  

In Fig.~\ref{SFOPT_1} we show the relationship between $\Delta \mathcal{F}_0 / \mathcal{F}^{SM}_0$ and $\xi_c$. 
The first order phase transition can only happen in the region with  $\Delta \mathcal{F}_0 / \mathcal{F}^{SM}_0 < 1.0$. 
This is consistent with the relationship found in~\cite{CPV}. 
Here we need to emphasize that even there is a strong correlation between $\Delta \mathcal{F}_0 / \mathcal{F}^{SM}_0$ and phase transition, it doesn't mean that the phase transition is decided by $\Delta \mathcal{F}_0 / \mathcal{F}^{SM}_0$ solely. 
In the region with $\Delta \mathcal{F}_0 / \mathcal{F}^{SM}_0 < 1.0$, the probability for $\xi_c > 0.0$ and $\xi_c > 1.0$ are 8.3\% and 0.8\%. 
Thus in order to get SFOPT, a certain level of parameter fine tuning is required. 
In our parameter space, $m_{H^\pm}$, $m_H$, and $m^2_{12}$ are closely related to $\Delta \mathcal{F}_0 / \mathcal{F}^{SM}_0 $, 
see plots in Fig.~\ref{SFOPT_1}. While $\tan\beta$ and $m_A$ are not so relevant to phase transition in our scenario. 
Furthermore, in Fig.~\ref{SFOPT_2} we project all the points on the planes of $m_H$ VS $m_{H^\pm}$, $\tan\beta$ VS $m^2_{12}$,
 and $\tan\beta$ VS $m_A$ with color mapped by $\xi_c$.
It can be seen that $\xi_c > 0.0$ and $\xi_c > 1.0$ constrain the planes of $m_H$ VS $m_{H^\pm}$ and $m^2_{12}$ VS 
$\tan\beta$ to very narrow regions,
 but the phase transition is not sensitive to $\tan\beta$ and $m_A$. 
The points with $\xi_c > 0.0$ and $m_A > 25$ GeV are excluded by the direct searches limits of LHC.

To conclude this section, SFOPT and the explanation of muon g-2 in L2HDM can happen in a small subset of 2HDM parameter space, 
where 14 GeV $<m_A<$ 25 GeV, 310 GeV $<m_H<$ 355 GeV, and 250 GeV $<m_{H^\pm}<$ 295 GeV.
We list detailed information of several benchmark points achieving the SFOPT and explaining the muon $g-2$ anomaly in Table \ref{tabpt}.

\begin{table}
\begin{footnotesize}
\begin{tabular}{| c | c | c | c | c |}
\hline
%%%%%%%%%%%%
 Benchmark points & $A$ & $B$ & $C$ & $D$\\
\hline
{$\sin(\beta-\alpha)$}       & 0.999    &  0.9989   & 0.9992    & 0.9987\\
{$\tan\beta$}                & 48.57    &  46.09    & 53.66     & 41.46\\
{$m_h$~(GeV)}                & 125.0    &  125.0    & 125.0     & 125.0\\
{$m_H$~(GeV)}                & 314.96   &  322.95   & 330.88    & 342.27\\
{$m_A$~(GeV)}                & 18.22    &  20.3     & 18.24     & 20.45\\
{$m_{H^\pm}$~(GeV)}          & 253.27   &  259.89   & 264.59    & 284.7\\
{$m_{12}^2$~(GeV$^2$)}      &~~2041.32~~&~~2261.78~~&~~2039.42~~&~~ 2823.33\\
{$\xi_c = v_c/T_c$} &1.015  &1.066  &1.117  &1.132    \\
\hline

\end{tabular}
\end{footnotesize}
\caption{Several benchmark points achieving the SFOPT.}
\label{tabpt}
\end{table}

\section{Conclusion}
The L2HDM can provide a simple explanation for the muon $g-2$ anomaly. 
We performed a scan over the parameter space of L2HDM to identify the ranges in
favor of the muon $g-2$ explanation after imposing various relevant theoretical and experimental constraints, 
especially the direct search limits from LHC and a SFOPT in the early universe. 
We found that the muon g-2 anomaly can be accommodated in the region of 32 $<\tan\beta<$ 80, 10 GeV $<m_A<$ 65 GeV,
260 GeV $<m_H<$ 620 GeV and 180 GeV $<m_{H^\pm}<$ 620 GeV after imposing the joint constraints from 
the theory, the precision electroweak data, the 125 GeV Higgs signal data, the LFU 
in $\tau$ and $Z$ decays, and the measurement of Br$(B_s \to \mu^+ \mu^-)$.
 The direct search limits from the LHC can give stringent constraints on $m_A$ and $\tan\beta$ 
for small $m_H$ and $m_{H^\pm}$: 10 GeV $<m_A<$ 44 GeV and 32 $<\tan\beta<$ 60.
The direct search limits from the $h\to AA$ channels at the LHC
can impose stringent upper limits on Br$(h\to AA)$.
Finally, we found that a SFOPT can be achievable in the region of
 14 GeV $<m_A<$ 25 GeV, 310 GeV $<m_H<$ 355 GeV, and 250 GeV $<m_{H^\pm}<$ 295 GeV 
while the muon $g-2$ anomaly is accommodated. 

\section*{Acknowledgment}
We appreciate the useful discussions with Fapeng Huang, Eibun Senaha and Wenlong Sang. 
This work was supported by the National Natural Science Foundation
of China under grant 11575152, 11675242, 11851303,
by the Natural Science Foundation of
Shandong province (ZR2017MA004 and ZR2017JL002),
by Peng-Huan-Wu Theoretical Physics Innovation Center (11747601),
by the CAS Center for Excellence in Particle Physics (CCEPP),
by the CAS Key Research Program of Frontier Sciences,
by a Key R\&D Program of Ministry of Science and Technology under number 2017YFA0402200-04, 
by IBS under the project code IBS-R018-D1,
and by the ARC Centre of Excellence for Particle Physics at the Tera-scale under the grant CE110001004.

\end{document}